\documentclass[journal]{IEEEtran}

\input{Qcircuit.tex }
\usepackage{amsmath}
\usepackage{graphicx}
\usepackage{subcaption}
\usepackage{float}
\usepackage{epstopdf}
\usepackage{multicol}
\usepackage{multirow}
\usepackage{rotating}
\usepackage{booktabs}
\usepackage[autostyle=false, style=english]{csquotes}
\MakeOuterQuote{"}
\usepackage{color}
\usepackage{algorithm}

\ifCLASSINFOpdf
\else
\fi

\begin{document}
\title{Quantum Circuit Designs of Integer Division Optimizing T-count and T-depth}

\author{\IEEEauthorblockN{Himanshu Thapliyal\IEEEauthorrefmark{1}, Edgard Mu\~{n}oz-Coreas\IEEEauthorrefmark{1}, T. S. S. Varun\IEEEauthorrefmark{1},   and Travis S. Humble\IEEEauthorrefmark{2}}

\thanks{\IEEEauthorrefmark{1} Himanshu Thapliyal, Edgard Mu\~{n}oz-Coreas, T. S. S. Varun are with the Department of Electrical and Computer Engineering, University of Kentucky, Lexington, KY, USA. \newline
Email :hthapliyal@uky.edu \newline
\IEEEauthorrefmark{2} Travis S. Humble is with the Quantum Computing Institute, Oak Ridge National Laboratory, TN, USA. \newline
}
}

\bstctlcite{IEEEexample:BSTcontrol}


%Telephone: (800) 555--1212, Fax: (888) 555--1212}


\maketitle

\begin{abstract}

Quantum circuits for mathematical functions such as division are necessary to use quantum computers for scientific computing.  Quantum circuits based on Clifford+T gates can easily be made fault-tolerant but the T gate is very costly to implement.  The small number of qubits available in existing quantum computers adds another constraint on quantum circuits.  As a result, reducing T-count and qubit cost have become important optimization goals.  The design of quantum circuits for integer division has caught the attention of researchers and designs have been proposed in the literature.  However, these designs suffer from excessive T gate and qubit costs.  Many of these designs also produce significant garbage output resulting in additional qubit and T gate costs to eliminate these outputs.  In this work, we  propose two quantum integer division circuits.  The first proposed quantum integer division circuit is based on the restoring division algorithm and the second proposed design implements the non-restoring division algorithm.  Both proposed designs are optimized in terms of T-count, T-depth and qubits.  Both proposed quantum circuit designs are based on (i) a quantum subtractor, (ii) a quantum adder-subtractor circuit, and (iii) a novel quantum conditional addition circuit.  Our proposed restoring division circuit achieves average T-count savings from $79.03 \%$ to $91.69 \%$ compared to the existing works.  Our proposed non-restoring division circuit achieves average T-count savings from $49.75 \%$ to $90.37 \%$ compared to the existing works.  Further, both our proposed designs have linear T-depth.

\end{abstract}


\IEEEpeerreviewmaketitle

\section{Introduction}

Among the emerging computing paradigms, quantum computing appears promising due to its applications in number 
theory, encryption, search and scientific computation \cite{Quipper} \cite{Hallgren2017encryption}.  Quantum circuits for integer arithmetic operations such as addition, subtraction, multiplication and division are required in the quantum circuit implementations of many quantum algorithms in these areas.   Quantum arithmetic circuits for division can be used in the circuit implementation of quantum algorithms such as those for computing shifted quadratic character problems, principal ideal problems and hidden shift problems \cite{Hallgren2017encryption} \cite{vandam2000quadraticcharacter} \cite{vandam2006hiddenshift}.  Quantum division circuits also reduce the resources needed in the circuit implementations of higher level functions such as calculating the greatest common divisor via the Euclidean algorithm.  An efficient quantum circuit for the Euclidean algorithm has use in quantum algorithms such as those for solving the shifted multiplicative character problem \cite{vandam2006hiddenshift}.  Thus, researchers have included dedicated libraries of basic quantum integer arithmetic functions such as division in quantum programming languages such as Quipper and $LIQUi\vert\rangle$ and in quantum computing design tools \cite{Quipper} \cite{LIQUi}.

Quantum computation can be performed on quantum circuits built from quantum gates.  Any constant inputs in the quantum circuit are called ancillae.  Garbage outputs are any outputs which exist in the quantum circuit to preserve one-to-one mapping but are neither one of the primary inputs nor a useful output.  The inputs regenerated at the outputs are not considered garbage outputs \cite{Fredkin}.  Ancillae and garbage outputs are circuit overhead that need to be minimized.

The fault-tolerant implementation of quantum circuits is gaining the attention of researchers because physical quantum computers are prone to noise errors \cite{Zhou-T}.  Fault-tolerant implementations of quantum gates and quantum error correcting codes can be used to overcome the limits imposed by noise errors in implementing quantum computing \cite{Maslov} \cite{PalerIOP}.  Recently, researchers have implemented quantum logic gates such as the Toffoli gate, Fredkin gate and quantum full adder with fault-tolerant implementations of the Clifford+T gates due to their demonstrated tolerance to noise errors \cite{Maslov}.  However, the increased tolerance to noise errors comes with the increased implementation overhead associated with the quantum T gate \cite{Gosset} \cite{Maslov} \cite{PalerIOP}.  Because of the increased cost to realize the T gate, T-count and T-depth are important performance measures for fault-tolerant quantum circuit design.

The design of quantum circuits for integer division is an active area of research.  Designs such as those proposed in \cite{khosropour2011quantum}, \cite{Jain2015quantumdivider} and \cite{Dibbo} present dividers that can be used in quantum computation.  The design in \cite{khosropour2011quantum} implements the restoring division algorithm and the designs in \cite{Jain2015quantumdivider} are based on the non-restoring division algorithm.  The design presented in the recent work in \cite{Dibbo} uses a novel division algorithm.  The design in \cite{khosropour2011quantum} has a significant overhead in terms of T gates because it depends on quantum gates that cannot be exactly constructed using Clifford+T gates.  The Clifford+T gate approximations for these gates are costly in terms of T-count \cite{Kliuchnikov}. Further, the T-count increases as the accuracy of the approximation of these gates is improved \cite{Kliuchnikov}.   In contrast the dividers in \cite{Jain2015quantumdivider} depend on quantum gates that can be exactly realized with Clifford+T gates.  At most 7 T gates are required to implement each logic gate in \cite{Jain2015quantumdivider}.  Thus, the designs in \cite{Jain2015quantumdivider} require significantly fewer T gates than the design presented in \cite{khosropour2011quantum}.  However, the designs in \cite{Jain2015quantumdivider} produce significant garbage output.  Thus, the dividers in \cite{Jain2015quantumdivider} will have additional ancillae and T gate overhead from removing these garbage outputs.  A recent design presented in \cite{Dibbo} also depends on quantum gates that can be exactly realized with Clifford+T gates.  However, the design methodology presented in \cite{Dibbo} generates a quantum circuit that suffers from significant T gate and qubit cost overhead.  In addition, the design in \cite{Dibbo} produces significant garbage output.  Thus, the divider in \cite{Dibbo} will have additional ancillae and T gate overhead from removing these garbage outputs.    
 
\textit{To address the shortcomings of the existing work this paper presents two designs for quantum circuit integer division based on Clifford+T gates. The first proposed quantum circuit is based on the restoring division algorithm and the second proposed quantum circuit is based on the non-restoring division algorithm.} Both proposed quantum integer division circuits are based on: (i) a quantum subtractor, (ii) a quantum adder-subtractor circuit, and (iii) a novel quantum conditional addition circuit. The preliminary version of this paper is available in \cite{Varun2016divider,mine2017divider}. The proposed quantum restoring division circuit has quadratic T-count, linear T-depth and requires $3 \cdot n$ qubits (where $n$ is the size of the inputs).  The proposed non-restoring division circuit has quadratic T-count, linear T-depth and requires $3 \cdot n - 1$ qubits (where $n$ is the size of the inputs).  

This paper is organized as follows: Section \ref{md_ref} discusses the Clifford+T gate set, background on resource cost measures, the quantum subtractor, quantum adder-subtractor circuit and the novel quantum conditional addition circuit.  The proposed quantum restoring integer division circuit is presented in Section \ref{INIS-res} while the comparison with the existing works is presented in Section \ref{Res-Cost}.  In Section \ref{INIS-nonres}, the design of the proposed quantum non-restoring integer division circuit is discussed and Section \ref{div-nrs-cost} illustrates comparison with the existing works.  

\section{Background}
\label{md_ref}

\subsection{Quantum Gates}

\begin{figure}[hbt]
\centering
\begin{subfigure}[hb]{2.5in}
\begin{center}
	\textsc{Clifford+T Gate Set}
\end{center}
\end{subfigure} \\ \begin{subfigure}[hb]{1in}
Hadamard Gate
\end{subfigure} \qquad \begin{subfigure}[hb]{.5in}
\[
 \Qcircuit @C=0.7em @R=0.5em @!R{
 & \gate{H} &  \qw & &   }
 \]
\end{subfigure} \qquad \begin{subfigure}[hb]{.75in}
\centering
  $ \frac{1}{\sqrt{2}}
  \begin{bmatrix}
    1 & 1  \\
    1 & -1 
  \end{bmatrix} $
\end{subfigure} \\ \begin{subfigure}[hb]{1in}

\end{subfigure} \qquad \begin{subfigure}[hb]{.5in}

\end{subfigure} \qquad \begin{subfigure}[hb]{.75in}

\end{subfigure} \\ \begin{subfigure}[hb]{1in}
T Gate
\end{subfigure} \qquad \begin{subfigure}[hb]{.5in}
\[
 \Qcircuit @C=0.7em @R=0.5em @!R{
 & \gate{T} &  \qw & &   }
 \]
\end{subfigure} \qquad \begin{subfigure}[hb]{.75in}
\centering
  $\begin{bmatrix}
    1 & 0  \\
    0 & e^{i \cdot \frac{\pi}{4}} 
  \end{bmatrix} $
\end{subfigure} \\ \begin{subfigure}[hb]{1in}

\end{subfigure} \qquad \begin{subfigure}[hb]{.5in}

\end{subfigure} \qquad \begin{subfigure}[hb]{.75in}

\end{subfigure} \\ \begin{subfigure}[hb]{1in}
\flushleft
Hermitian of T Gate
\end{subfigure} \qquad \begin{subfigure}[hb]{.5in}
\[
 \Qcircuit @C=0.7em @R=0.5em @!R{
 & \gate{T^{\dag}} &  \qw & &   }
 \]
\end{subfigure} \qquad \begin{subfigure}[hb]{.75in}
\centering
    $\begin{bmatrix}
    1 & 0  \\
    0 & e^{-i \cdot \frac{\pi}{4}} 
  \end{bmatrix} $
\end{subfigure} \\ \begin{subfigure}[hb]{1in}

\end{subfigure} \qquad \begin{subfigure}[hb]{.5in}

\end{subfigure} \qquad \begin{subfigure}[hb]{.75in}

\end{subfigure} \\ \begin{subfigure}[hb]{1in}
Phase Gate
\end{subfigure} \qquad \begin{subfigure}[hb]{.5in}
 \[
 \Qcircuit @C=0.7em @R=0.5em @!R{
 & \gate{S} &  \qw & &   }
 \]
\end{subfigure} \qquad \begin{subfigure}[hb]{.75in}
\centering
  $\begin{bmatrix}
    1 & 0  \\
    0 & i 
  \end{bmatrix} $
\end{subfigure} \\ \begin{subfigure}[hb]{1in}

\end{subfigure} \qquad \begin{subfigure}[hb]{.5in}

\end{subfigure} \qquad \begin{subfigure}[hb]{.75in}

\end{subfigure} \\ \begin{subfigure}[hb]{1in}
\flushleft
Hermitian of Phase Gate
\end{subfigure} \qquad \begin{subfigure}[hb]{.5in}
 \[
 \Qcircuit @C=0.7em @R=0.5em @!R{
 & \gate{S^{\dag}} &  \qw & &   }
 \]
\end{subfigure} \qquad \begin{subfigure}[hb]{.75in}
\centering
  $\begin{bmatrix}
    1 & 0  \\
    0 & -i 
  \end{bmatrix} $
\end{subfigure} \\ \begin{subfigure}[hb]{1in}

\end{subfigure} \qquad \begin{subfigure}[hb]{.5in}

\end{subfigure} \qquad \begin{subfigure}[hb]{.75in}

\end{subfigure} \\ \begin{subfigure}[hb]{1in}
NOT Gate
\end{subfigure} \qquad \begin{subfigure}[hb]{.5in}
  \[
 \Qcircuit @C=0.7em @R=0.5em @!R{
 & \targ &  \qw & &   }
 \]
\end{subfigure} \qquad \begin{subfigure}[hb]{.75in}
\centering
 $\begin{bmatrix}
    0 & 1  \\
    1 & 0 
  \end{bmatrix}$
\end{subfigure} \\ \begin{subfigure}[hb]{1in}

\end{subfigure} \qquad \begin{subfigure}[hb]{.5in}

\end{subfigure} \qquad \begin{subfigure}[hb]{.75in}

\end{subfigure} \\ \begin{subfigure}[hb]{1in}
Feynman (CNOT) Gate
\end{subfigure} \qquad \begin{subfigure}[hb]{.5in}
  \[
 \Qcircuit @C=0.7em @R=0.5em @!R{
 & \ctrl{1} &  \qw & & \\ 
 & \targ &  \qw & & }
 \]
\end{subfigure} \qquad \begin{subfigure}[hb]{.75in}
\centering
  $\begin{bmatrix}
    1 & 0 & 0 & 0 \\
    0 & 1 & 0 & 0\\
    0 & 0 & 0 & 1 \\
    0 & 0 & 1 & 0 \\
  \end{bmatrix} $ 
\end{subfigure}
\caption{The quantum gate set used in this work.}
\label{Clifford table}
\end{figure}

Fault-tolerant implementation of quantum circuits is gaining the attention of researchers because physical quantum computers are prone to noise errors \cite{Zhou-T}.  Recently, researchers have implemented quantum logic gates and circuits with fault-tolerant implementations of the Clifford+T gate set due to its demonstrated tolerance to noise errors \cite{Maslov}.  Figure \ref{Clifford table} presents the gates that make up the Clifford+T quantum gate family. Evaluating quantum circuit performance in terms of T-count and T-depth is of interest to researchers because the fault-tolerant implementation of the T gate is significantly more costly than the fault-tolerant implementation costs of the other Clifford+T gates \cite{Gosset}. The number of qubits in a quantum circuit is a resource measure of interest because of the limited number of qubits available on existing quantum computers \cite{IBM_quantum} \cite{Song2017computersaresmall}.  We now define the T-count, T-depth and qubit cost resource measures.

\begin{itemize} 
\item T-count: T-count is the total number of T gates used in the quantum circuit.
\item T-depth: T-depth is the number of T gate layers in the circuit, where a layer consists of quantum operations that can be performed simultaneously.
\item Qubit cost: Qubit cost is the total number of qubits required to design the quantum circuit.
\end{itemize}

\subsection{Design of Quantum Subtractor}

The subtractor circuit takes two $n$ bit inputs $a$ and $b$.  At the end of computation, the input $a$ emerges unchanged and the input $b$ is transformed to the difference of $b$ from $a$.  The quantum subtractor calculates $(\overline{\bar{b}+a})$ which is equivalent to $b-a$ \cite{Thapliyal2016addsub}. A quantum ripple carry adder is used to realize the quantum subtractor circuit. We use the quantum ripple carry adder proposed in \cite{thapliyal2013design} for developing the quantum subtractor circuit. We chose the ripple carry adder proposed in \cite{thapliyal2013design} because this adder has been shown in the literature to be efficient in terms of gates and has a T-depth that is constant and independent of the circuit size $n$. We determined that this quantum subtractor will have a T-count of $14 \cdot n - 14$ and a T-depth of $10$.  Thus, the quantum subtraction circuit used in our proposed dividers will have a T-count of order $\mathcal{O}(n)$ and a T-depth of order $\mathcal{O}(1)$.

\subsection{ Design of Quantum Adder-Subtractor (\textit{Add-Sub}) Circuit}   %resume here

The quantum adder-subtractor (\textit{Add-Sub}) circuit takes two $n$ bit inputs $a$ and $b$ and a single one bit input $ctrl$.  Operation of the quantum \textit{Add-Sub} circuit is conditioned on the value of $ctrl$.  When $ctrl$ is high, the circuit calculates  $b-a$.  When the $ctrl$ input is low, the circuit calculates $b+a$. The quantum \textit{Add-Sub} calculates $(\overline{\bar{b}+a})$ when $ctrl$ is high.  The expression $(\overline{\bar{b}+a})$ is equivalent to $b-a$.  The quantum \textit{Add-Sub} circuit is based on the design presented in \cite{Thapliyal2016addsub} and uses the ripple carry adder in \cite{thapliyal2013design}. We determined that this quantum \textit{Add-Sub} circuit will have a T-count of $14 \cdot n - 14$ and a T-depth of $10$.  Thus, the quantum \textit{Add-Sub} circuit used in our proposed dividers will have a T-count of order $\mathcal{O}(n)$ and a T-depth of order $\mathcal{O}(1)$. 

 \subsection{Design of Quantum Conditional Addition \textit{Ctrl-Add} Circuit}

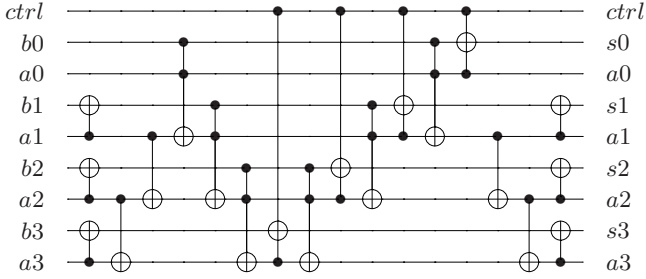
\begin{figure}[htbp]
	\centering
	\small
	
	\[
	\Qcircuit @C=0.5em @R=0.5em @!R{
		\lstick{ctrl}&	&\qw			&\qw			&\qw		
		&\qw			&\qw			&\qw	
		&\ctrl{7}		&\qw			
		&\ctrl{5}		&\qw			&\ctrl{3}	
		&\qw			&\ctrl{1}		&\qw	
		&\qw			&\qw			
		&\qw		
		&\rstick{ctrl}\\
		\lstick{b0}&	&\qw			&\qw			&\qw		
		&\ctrl{3}		&\qw			&\qw	
		&\qw			&\qw			
		&\qw			&\qw			&\qw		
		&\ctrl{3}		&\targ		&\qw		
		&\qw			&\qw			&\qw	
		&\rstick{s0}\\
		\lstick{a0}&	&\qw			&\qw			&\qw		
		&\ctrl{2}		&\qw			&\qw	
		&\qw			&\qw			
		&\qw			&\qw			&\qw		
		&\ctrl{2}		&\ctrl{-1}		&\qw	
		&\qw			&\qw			
		&\qw			&\rstick{a0}\\
		\lstick{b1}&	&\targ		&\qw			&\qw			
		&\qw			&\ctrl{3}		&\qw		
		&\qw			&\qw			&\qw	
		&\ctrl{3}		&\targ		&\qw	
		&\qw			&\qw			
		&\qw			&\targ		&\qw			
				&\rstick{s1}\\
		\lstick{a1}&	&\ctrl{-1}	&\qw			&\ctrl{2}	
		&\targ		&\ctrl{2}		&\qw		
		&\qw			&\qw			&\qw	
		&\ctrl{2}		&\ctrl{-1}		
		&\targ		&\qw			&\ctrl{2}		
		&\qw			&\ctrl{-1}		&\qw		
				&\rstick
		{a1}\\
		\lstick{b2}&	&\targ		&\qw	
		&\qw			
		&\qw			&\qw			&\ctrl{3}	
		&\qw			&\ctrl{3}		&
		\targ		&\qw			&\qw			
		&\qw			&\qw			&\qw		
		&\qw			&\targ				&\qw		&
		\rstick{s2}\\
		\lstick{a2}&	&\ctrl{-1}		
		&\ctrl{2}		&\targ		
		&\qw			&\targ		&\ctrl{2}		
		&\qw			&\ctrl{2}		&\ctrl{-1}	
		&\targ		&\qw			&\qw		
		&\qw			&\targ		&\ctrl{2}	
					&\ctrl{-1}	
		&\qw		&\rstick{a2}\\
		\lstick{b3}&	&\targ		&\qw			&\qw			
		&\qw			&\qw			&\qw		
		&\targ		&\qw			&\qw		
		&\qw			&\qw			&\qw	
		&\qw			&\qw			
		&\qw				
		&\targ		&\qw		&\rstick{s3}\\
		\lstick{a3}&	&\ctrl{-1}		&\targ		&\qw			
		&\qw			&\qw			&\targ		
		&\ctrl{-1}		&\targ		&\qw			
		&\qw			&\qw			&\qw		
		&\qw			&\qw			&
		\targ			
		&\ctrl{-1}		&\qw		&\rstick{a3}
	}
	\]
	
	\caption{Circuit design of quantum Conditional Addition (\textit{Ctrl-Add}) circuit }
	\label{add nop}
\end{figure}

\textit{The quantum \textit{Ctrl-Add} circuit in this work is a modified version of the \textit{Ctrl-Add} circuit proposed in \cite{edgard2017multiplier}.} Operation of the quantum \textit{Ctrl-Add} circuit is conditioned on the value of $ctrl$.  When $ctrl$ is high, the circuit calculates $b + a$.  When $ctrl$ is low, the circuit performs no computation.  An illustrative example is shown in Figure \ref{add nop} for two 4 qubit operands $a_0 \dots a_3$ and $b_0 \dots b_3$.  We are able to reduce the amount of qubits and quantum gates required because we do not need the carry out qubit in the proposed integer dividers. 

We determined that our quantum \textit{Ctrl-Add} circuit will have a T-count of $21 \cdot n - 14$ and a T-depth of $2 \cdot n$.  Thus, the quantum \textit{Ctrl-Add} circuit used in our proposed dividers will have a T-count of order $\mathcal{O}(n)$ and a T-depth of order $\mathcal{O}(n)$. 

%%%%%%%%%%%%%%%%%%%%%%%

\section{Proposed Design of Restoring Quantum Integer Division Circuit}
\label{INIS-res}

We now present our proposed restoring quantum integer division circuit.  The proposed design produces no garbage output and has lower T-count and qubit costs compared to the existing works.  The quantum circuits used in our proposed quantum restoring division circuit are (i) the quantum subtractor and (ii) the quantum \textit{Ctrl-Add} circuit.  Our proposed quantum restoring divider saves T gates by not doing computation in the QFT domain.  We also base our design on the T gate efficient quantum subtractor and the novel quantum \textit{Ctrl-Add} circuit presented in Section \ref{md_ref}.  The modules used in our quantum circuit do not produce garbage outputs and restore inputs to their original values.  Thus, we are able to save qubits and T gates by placing these quantum circuits such that our proposed quantum restoring division circuit will produce no garbage outputs.  

This proposed quantum integer division circuit calculates division by implementing the restoring division algorithm.  The restoring division algorithm is illustrated in Figure \ref{sqrt-table:rsalg}.  Prior research has demonstrated the correctness of the restoring division algorithm through functionally correct circuit implementations such as those in \cite{khosropour2011quantum}.

Consider the division of two $n$ bit $2$'s complement positive binary numbers $a$ and $b$.   Let $\ket{B}$ be a $n$ bit quantum register that is initialized to the value $b$, let $\ket{Q}$ be a $n$ bit quantum register that is initialized with the value $a$ and let $\ket{R}$ be a $n$ bit quantum register initialized to $0$.  At the end of computation, the quantum register $\ket{B}$ will be restored to the value $b$ while the quantum register $\ket{R}$ will have the remainder of the division of $a$ by $b$.  At the end of computation, the quantum register $\ket{Q}$ will have the quotient of the division of $a$ by $b$.

\begin{figure}[thbp]

\begin{tabular}{ll}
\\ \midrule
\multicolumn{2}{l}{\textbf{Algorithm 1:} Restoring division algorithm }\\ \toprule
\multicolumn{2}{l}{ \textbf{Function} Restoring($a,b$) }\\ 
\multicolumn{2}{l}{ Requirements: $a$ and $b$ are positive and $2$'s complement. }\\
\multicolumn{2}{l}{ \qquad //Takes 2 $n$ bit values $a$ and $b$ as input. }\\
\multicolumn{2}{l}{ \qquad //Returns the quotient as an $n$ bit number $Q$ and }\\
\multicolumn{2}{l}{ \qquad //the remainder from the division as an $n$ bit }\\
\multicolumn{2}{l}{ \qquad //number $R$. }\\  
&  \\
1 & $R = 0^{n}$; // Where $0^{n}$ are $n$ zeros. \\
2 & $Q = a_{n-1} a_{n-2} \cdots a_1 a_0$ \\
3 & \qquad // $a_{n-1}$ is the most significant bit of $a$. \\
4 & \\
5 & \textbf{For} $i = 1 \text{ to } n-1$ \\ 
6 & \qquad $Y = R_{n-1-i} R_{n-3-i} \cdots R_1 R_0 Q_{n-1} \cdots Q_{n-i}$ \\
7 & \qquad // Where $R_{n-1-i}$ is the most significant bit of $Y$. \\
8 & \qquad $Y = Y - b$ \\
9 & \qquad \textbf{If} $(Y < 0)$ \\
10 & \qquad \qquad $Y = Y + b$ \\
11 & \qquad \textbf{End} \\
12 & \qquad $R_{n-i} = \overline{R_{n-1-i}}$ \\
13 & \textbf{End} \\
14 & \\
15 &  $Q = Q - b$ \\
16 & \textbf{If} $(Q < 0)$ \\
17 & \qquad $Q = Q + b$; \\
18 & \textbf{End} \\
19 & $R_{0} = \overline{Q_{n-1}}$ \\
20 & \\ 
21 & \textbf{Return} $Q, R$ \\ \bottomrule
\end{tabular}

\caption{The restoring division algorithm. }
\label{sqrt-table:rsalg}
\end{figure}

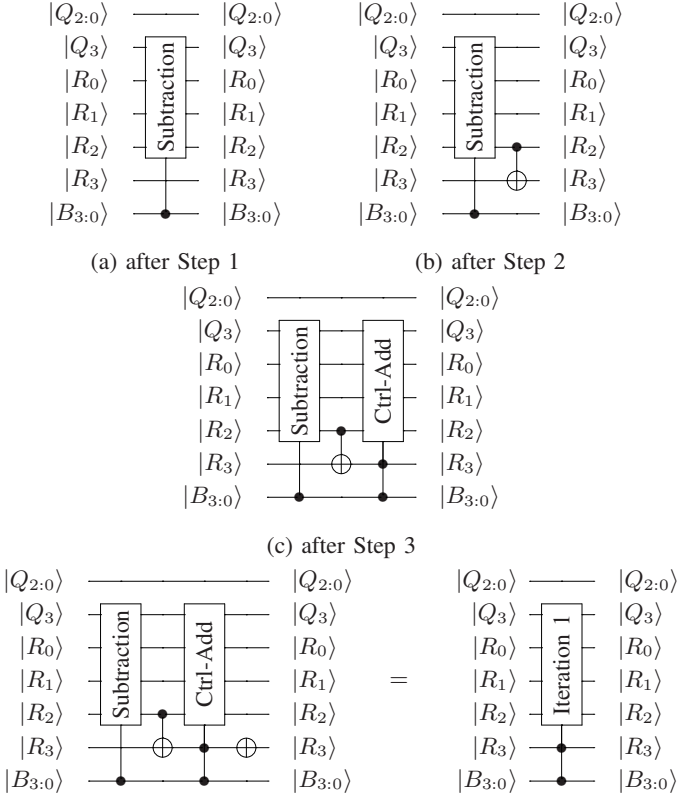
\begin{figure}[h]
\centering
\begin{subfigure}[th]{1.25in}
\small
\[
\Qcircuit @C=0.5em @R=0.5em @!R{
\lstick{\ket{Q_{2:0}}}&	&\qw			&\qw		&\rstick{\ket{Q_{2:0}}}\\
\lstick{\ket{Q_{3}}}&	&\multigate{3}{\begin{sideways} Subtraction \end{sideways}}		&\qw		&\rstick{\ket{Q_{3}}}\\
\lstick{\ket{R_{0}}}&	&\ghost{\begin{sideways} Subtraction \end{sideways}}		&\qw			&\rstick{\ket{R_{0}}}\\
\lstick{\ket{R_{1}}}&	&\ghost{\begin{sideways} Subtraction \end{sideways}}		&\qw			&\rstick{\ket{R_{1}}}\\
\lstick{\ket{R_{2}}}&	&\ghost{\begin{sideways} Subtraction \end{sideways}}		&\qw			&\rstick{\ket{R_{2}}}\\
\lstick{\ket{R_{3}}}&	&\qw	&\qw		&\rstick{\ket{R_{3}}}\\
\lstick{\ket{B_{3:0}}}&	&\ctrl{-2}		&\qw			&\rstick{\ket{B_{3:0}}}
}
\]
\caption{after Step 1}
\label{s1res}
\end{subfigure} \qquad \begin{subfigure}[th]{1.5in}
\small
\[
\Qcircuit @C=0.5em @R=0.5em @!R{
\lstick{\ket{Q_{2:0}}}&	&\qw			&\qw		&\qw &\rstick{\ket{Q_{2:0}}}\\
\lstick{\ket{Q_{3}}}&	&\multigate{3}{\begin{sideways} Subtraction \end{sideways}}		&\qw &\qw		&\rstick{\ket{Q_{3}}}\\
\lstick{\ket{R_{0}}}&	&\ghost{\begin{sideways} Subtraction \end{sideways}}		&\qw &\qw			&\rstick{\ket{R_{0}}}\\
\lstick{\ket{R_{1}}}&	&\ghost{\begin{sideways} Subtraction \end{sideways}}		&\qw &\qw			&\rstick{\ket{R_{1}}}\\
\lstick{\ket{R_{2}}}&	&\ghost{\begin{sideways} Subtraction \end{sideways}}		&\ctrl{1} &\qw			&\rstick{\ket{R_{2}}}\\
\lstick{\ket{R_{3}}}&	&\qw	&\targ &\qw		&\rstick{\ket{R_{3}}}\\
\lstick{\ket{B_{3:0}}}&	&\ctrl{-2}		&\qw		&\qw	&\rstick{\ket{B_{3:0}}}
}
\]
\caption{after Step 2}
\label{s2res}
\end{subfigure}
\\
\begin{subfigure}[hb]{3.5in}
\small
\[
\Qcircuit @C=0.5em @R=0.5em @!R{
\lstick{\ket{Q_{2:0}}}&	&\qw			&\qw		&\qw &\qw &\rstick{\ket{Q_{2:0}}}\\
\lstick{\ket{Q_{3}}}&	&\multigate{3}{\begin{sideways} Subtraction \end{sideways}}		&\qw &\multigate{3}{\begin{sideways} Ctrl-Add \end{sideways}} &\qw		&\rstick{\ket{Q_{3}}}\\
\lstick{\ket{R_{0}}}&	&\ghost{\begin{sideways} Subtraction \end{sideways}}		&\qw &\ghost{\begin{sideways} Ctrl-Add \end{sideways}} &\qw			&\rstick{\ket{R_{0}}}\\
\lstick{\ket{R_{1}}}&	&\ghost{\begin{sideways} Subtraction \end{sideways}}		&\qw &\ghost{\begin{sideways} Ctrl-Add \end{sideways}} &\qw			&\rstick{\ket{R_{1}}}\\
\lstick{\ket{R_{2}}}&	&\ghost{\begin{sideways} Subtraction \end{sideways}}		&\ctrl{1} &\ghost{\begin{sideways} Ctrl-Add \end{sideways}} &\qw			&\rstick{\ket{R_{2}}}\\
\lstick{\ket{R_{3}}}&	&\qw	&\targ &\ctrl{-1} &\qw		&\rstick{\ket{R_{3}}}\\
\lstick{\ket{B_{3:0}}}&	&\ctrl{-2}		&\qw	&\ctrl{-2}	&\qw	&\rstick{\ket{B_{3:0}}}
}
\]
\caption{after Step 3}
\label{s3res}
\end{subfigure}
\\
\begin{subfigure}[hb]{3.5in}
\small
\[
\Qcircuit @C=0.5em @R=0.5em @!R{
\lstick{\ket{Q_{2:0}}}&	&\qw			&\qw		&\qw &\qw &\qw &\rstick{\ket{Q_{2:0}}} & & & & & & & & & & & & & & & & & & & & & \lstick{\ket{Q_{2:0}}}& \qw & \qw &\rstick{\ket{Q_{2:0}}} \\
\lstick{\ket{Q_{3}}}&	&\multigate{3}{\begin{sideways} Subtraction \end{sideways}}		&\qw &\multigate{3}{\begin{sideways} Ctrl-Add \end{sideways}} &\qw &\qw		&\rstick{\ket{Q_{3}}} & & & & & & & & & & & & & & & & & & & & & \lstick{\ket{Q_{3}}}& \multigate{3}{\begin{sideways} Iteration 1 \end{sideways}} & \qw &\rstick{\ket{Q_{3}}} \\
\lstick{\ket{R_{0}}}&	&\ghost{\begin{sideways} Subtraction \end{sideways}}		&\qw &\ghost{\begin{sideways} Ctrl-Add \end{sideways}} &\qw		&\qw	&\rstick{\ket{R_{0}}} & & & & & & & & & & & & & & & & & & & & & \lstick{\ket{R_{0}}}& \ghost{\begin{sideways} Iteration 1 \end{sideways}} & \qw &\rstick{\ket{R_{0}}} \\
\lstick{\ket{R_{1}}}&	&\ghost{\begin{sideways} Subtraction \end{sideways}}		&\qw &\ghost{\begin{sideways} Ctrl-Add \end{sideways}} &\qw		&\qw	&\rstick{\ket{R_{1}}} & & & & &  & & &  & & = & & &  & & & & & & & & \lstick{\ket{R_{1}}}& \ghost{\begin{sideways} Iteration 1 \end{sideways}} & \qw &\rstick{\ket{R_{1}}} \\
\lstick{\ket{R_{2}}}&	&\ghost{\begin{sideways} Subtraction \end{sideways}}		&\ctrl{1} &\ghost{\begin{sideways} Ctrl-Add \end{sideways}} &\qw	&\qw		&\rstick{\ket{R_{2}}} & & & & & & & & & & & & & & & & & & & & & \lstick{\ket{R_{2}}}& \ghost{\begin{sideways} Iteration 1 \end{sideways}} & \qw &\rstick{\ket{R_{2}}} \\
\lstick{\ket{R_{3}}}&	&\qw	&\targ &\ctrl{-1} &\targ &\qw		 &\rstick{\ket{R_{3}}} & & & & & & & & & & & & & & & & & & & & & \lstick{\ket{R_{3}}}& \ctrl{-1} & \qw &\rstick{\ket{R_{3}}} \\
\lstick{\ket{B_{3:0}}}&	&\ctrl{-2}		&\qw	&\ctrl{-2}	&\qw &\qw	&\rstick{\ket{B_{3:0}}} & & & & & & & & & & & & & & & & & & & & & \lstick{\ket{B_{3:0}}}& \ctrl{-2} & \qw &\rstick{\ket{B_{3:0}}}
}
\]
\caption{after Step 4.  Quantum circuit and equivalent graphical representation are shown.}
\label{s4res}
\end{subfigure}
\caption{Circuit generation of the proposed quantum restoring division circuit for the case of $i = 1$ (illustrated with a 4 bit divider). }
\label{restoring 1 iter}
\end{figure}

%%%%%%%%%%%%%%%%%%%%%%%%%%

The proposed methodology is generic in nature and can design a quantum restoring integer division circuit of any size.  The steps of the proposed methodology are presented along with an illustrative example of the proposed quantum restoring integer division circuit for the division of two $4$ bit numbers $a_0 \dots a_3$ and $b_0 \dots b_3$ shown in Figure \ref{restoring 1 iter}.  The proposed methodology is repeated $n$ times.  A quantum circuit is generated for each step of the design.   

%%%%%%%%%%%%%%%%%%%

\subsection{Steps of the Proposed Design Methodology}

The following steps of the proposed methodology are repeated $n$ times.  Starting with the first $n-1$ iterations. 

For $i = 1:1:n-1$:

\begin{itemize}

\item Step 1: This step executes line 6 and line 8 of Algorithm 1 in quantum hardware.  Step 1 is shown for a 4 bit restoring divider in Figure \ref{s1res}.  This step has the following two sub-steps:    

\begin{itemize}

\item Sub-step 1: Treat the locations $\ket{R_{n-1-i}}$ through $\ket{R_0}$ of quantum register $\ket{R}$ and locations $\ket{Q_{n-1}}$ through $\ket{Q_{n-i}}$ of quantum register $\ket{Q}$
 as one $n$ qubit combined quantum register $\ket{Y}$ such that the values at locations $\ket{Q_{n-1}}$ through $\ket{Q_{n-i}}$ will occupy locations $\ket{Y_{i-1}}$ through $\ket{Y_0}$ and locations $\ket{R_{n-1-i}}$ through $\ket{R_0}$ will occupy locations $\ket{Y_{n-1}}$ through $\ket{Y_{i}}$.  
 
\item Sub-step 2: Apply quantum register $\ket{B}$ and quantum register $\ket{Y}$ to the quantum subtraction circuit so that, at the end of computation, $\ket{B}$ is unchanged and $\ket{Y}$ has the result of computation.  

\end{itemize}

\item Step 2: This step prepares register location $\ket{R_{n-i}}$ for use in subsequent steps and partially completes the execution of line 12 of Algorithm 1 in quantum hardware.  Step 2 is shown for a 4 bit restoring divider in Figure \ref{s2res}.  At quantum register locations $\ket{R_{n-i}}$ and $\ket{Y_{n-1}}$, apply a CNOT gate such that quantum register location $\ket{Y_{n-1}}$ is unchanged and quantum register location $\ket{R_{n-i}}$ is transformed to $\ket{R_{n-i} \oplus Y_{n-1}}$.  

\item Step 3: Step 3 implements lines 9 through 11 of Algorithm 1 in quantum hardware.  Step 3 is shown for a 4 bit restoring divider in Figure \ref{s3res}.  This step has the following two sub-steps:

\begin {itemize}

\item Sub-step 1:  Apply the quantum registers $\ket{B}$ and $\ket{Y}$ to a quantum \textit{Ctrl-Add} circuit with no input carry such that the quantum register $\ket{B}$ will be unchanged and quantum register $\ket{Y}$ will contain the result of computation.

\item Sub-step 2: Apply quantum register location $\ket{R_{n-i}}$ to the quantum \textit{Ctrl-Add} circuit such that the operation of the \textit{Ctrl-Add} circuit will be conditioned on the value at location $\ket{R_{n-i}}$.  Quantum register location $\ket{R_{n-i}}$ is unchanged.

\end {itemize}

After this step, 

\item Step 4: This step completes the execution of line 12 of Algorithm 1.  Step 4 is shown for a 4 bit restoring divider in Figure \ref{s4res}.  At quantum register location $\ket{R_{n-i}}$ apply a quantum NOT gate.  Quantum register location $\ket{R_{n-i}}$ now has the remainder bit $r_{n-i}$ of the division of $a$ by $b$. 
    
\end{itemize}

\begin{itemize}

\item[] The final iteration has the following steps:

\item Step 1: This step executes line 15 of Algorithm 1 in quantum hardware.  Step 1 is shown for a 4 bit restoring divider in Figure \ref{s1res}.  Apply quantum register $\ket{B}$ and quantum register $\ket{Q}$ to the quantum subtraction circuit so that, at the end of computation, $\ket{B}$ is unchanged and $\ket{Q}$ has the result of computation.  

\item Step 2: This step prepares register location $\ket{R_{0}}$ for use in subsequent steps and partially completes the execution of line 19 of Algorithm 1.  Step 2 is shown for a 4 bit restoring divider in Figure \ref{s2res}.  At quantum register locations $\ket{R_{0}}$ and $\ket{Q_{n-1}}$, apply a CNOT gate such that quantum register location $\ket{Q_{n-1}}$ is unchanged and quantum register location $\ket{R_{0}}$ shall be transformed to the value $\ket{R_{0} \oplus Q_{n-1}}$.     

\item Step 3: Steps 3 implements lines 16 through 18 of Algorithm 1 in quantum hardware.  Step 3 is shown for a 4 bit restoring divider in Figure \ref{s3res}.  Step 3 has the following two sub-steps:

\begin {itemize}

\item Sub-step 1:  Apply the quantum registers $\ket{B}$ and $\ket{Q}$ to a quantum \textit{Ctrl-Add} circuit with no input carry such that the quantum register $\ket{B}$ will be unchanged and quantum register $\ket{Q}$ will contain the result of computation.

\item Sub-step 2: Apply quantum register location $\ket{R_{0}}$ to the quantum \textit{Ctrl-Add} circuit such that the operation of the \textit{Ctrl-Add} circuit will be conditioned on the value at location $\ket{R_{0}}$.  Quantum register location $\ket{R_{0}}$ is unchanged.

\end {itemize}

After this step, quantum register $\ket{Q}$ will have the quotient of the division of $a$ and $b$.

\item Step 4: This step completes the execution of line 19 of Algorithm 1.  Step 4 is shown for a 4 bit restoring divider in Figure \ref{s4res}.  At quantum register location $\ket{R_{0}}$ apply a quantum NOT gate.  Quantum register location $\ket{R_{0}}$ now has the remainder bit $r_{0}$ of the division of $a$ by $b$. After this step, quantum register $\ket{R}$ now has the remainder of the division of $a$ and $b$.    

\end{itemize}

\begin{figure}[tbhp]

\centering

\[
\Qcircuit @C=1em @R=1em @!R{
\lstick{\ket{Q_0}}& \qw &\qw &\qw &\multigate{3}{\begin{sideways} Iteration 4 \end{sideways}} &\qw & \rstick{\ket{Q_0}} \\
\lstick{\ket{Q_1}}& \qw &\qw &\multigate{3}{\begin{sideways} Iteration 3 \end{sideways}} & \ghost{\begin{sideways} Iteration 4 \end{sideways}} &\qw & \rstick{\ket{Q_1}} \\
\lstick{\ket{Q_2}}& \qw &\multigate{3}{\begin{sideways} Iteration 2 \end{sideways}} & \ghost{\begin{sideways} Iteration 3 \end{sideways}} & \ghost{\begin{sideways} Iteration 4 \end{sideways}} &\qw & \rstick{\ket{Q_2}}\\
\lstick{\ket{Q_3}}& \multigate{3}{\begin{sideways} Iteration 1 \end{sideways}} & \ghost{\begin{sideways} Iteration 2 \end{sideways}} & \ghost{\begin{sideways} Iteration 3 \end{sideways}} & \ghost{\begin{sideways} Iteration 4 \end{sideways}} &\qw & \rstick{\ket{Q_3}} \\
\lstick{\ket{R_0}}& \ghost{\begin{sideways} Iteration 1 \end{sideways}} & \ghost{\begin{sideways} Iteration 2 \end{sideways}} & \ghost{\begin{sideways} Iteration 3 \end{sideways}} &\ctrl{-1} &\qw & \rstick{\ket{R_0}} \\
\lstick{\ket{R_1}}& \ghost{\begin{sideways} Iteration 1 \end{sideways}} & \ghost{\begin{sideways} Iteration 2 \end{sideways}} &\ctrl{-1} &\qw &\qw & \rstick{\ket{R_1}} \\
\lstick{\ket{R_2}}& \ghost{\begin{sideways} Iteration 1 \end{sideways}} &\ctrl{-1}  &\qw &\qw &\qw & \rstick{\ket{R_2}} \\
\lstick{\ket{R_3}}& \ctrl{-1} &\qw &\qw &\qw &\qw & \rstick{\ket{R_3}}\\
\lstick{\ket{B_{3:0}}}& \ctrl{-2} &\ctrl{-2} &\ctrl{-4} &\ctrl{-4} &\qw & \rstick{\ket{B_{3:0}}}
}
\]

\caption{Complete proposed quantum restoring integer divider circuit (after all $4$ iterations)}
\label{restoring n iter}
\end{figure}
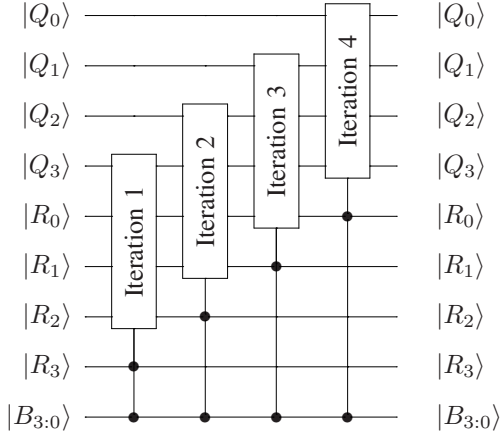

Figure \ref{restoring n iter} shows the complete proposed quantum restoring division circuit for the division of two $4$ bit numbers $a_0 \dots a_3$ and $b_0 \dots b_3$.

\section{Cost Analysis of the Proposed Restoring Division Circuit}
\label{Res-Cost}

\subsection{T-Count Analysis}

The T-count of the proposed quantum integer division circuit is illustrated shortly for each step of the proposed design methodology.  The steps are iterated $n$ times.  

\begin{itemize}

\item The T-count for Step 1 is $14 \cdot n - 14$.  We use a quantum subtraction circuit of T-count $14 \cdot n - 14$ in this step.

\item Steps 2 does not require T gates.

\item The T-count for Step 3 is $21 \cdot n - 14$.  We use a quantum \textit{Ctrl-Add} circuit of T-count $21 \cdot n - 14$ in this step.

\item Steps 4 does not require T gates.

\end{itemize}

We determine the T-count for a single iteration of the proposed design by summing the T-count for each step in the methodology as shown below:

\begin{equation}
14 \cdot n - 14 + 21 \cdot n - 14
\label{nrsdiv-EQ:10}
\end{equation}

This expression can be simplified to the following:

\begin{equation}
35 \cdot n -28
\label{nrsdiv-EQ:11}
\end{equation}

The steps in the proposed methodology are iterated $n$ times.  Thus the T-count for the proposed restoring division circuit is $n \cdot (35 \cdot n - 18)$ which simplifies to the expression shown below:

\begin{equation}
35 \cdot n^2 -28 \cdot n
\label{nrsdiv-EQ:12}
\end{equation}

\subsection{T-Depth Analysis}

The T-depth of the proposed quantum integer division circuit is illustrated shortly for each step of the proposed design methodology.  Our proposed design is based on T-depth efficient designs of quantum subtraction circuits and quantum \textit{Ctrl-Add} circuits.  We determined that garbageless and T gate optimized quantum subtraction circuits in the literature such as the design in \cite{Thapliyal2016addsub} have a T-depth that is constant and independent of the circuit size $n$.  Thus, these subtraction circuits have T-depth of order $\mathcal{O}(1)$.  We determined as well that \textit{Ctrl-Add} circuits in the literature such as the design in \cite{edgard2017multiplier} scale as a function of circuit size $n$.  Thus, these \textit{Ctrl-Add} circuits have a T-depth of order $\mathcal{O}(n)$.  

The T-depth of the proposed quantum integer division circuit is illustrated shortly for each step of the proposed design methodology.  The steps are iterated $n$ times.  

\begin{itemize}

\item Step 1 has a constant T-depth of $10$.  This T-depth is seen by locations $\ket{B_{1}}$ through $\ket{B_{n-2}}$ of quantum register $\ket{B}$.  We use a quantum subtraction circuit of constant T-depth $10$ in this step (where T-depth is independent of circuit size $n$). 

\item Step 2 does not require T gates.

\item Step 3 has a T-depth of $2 \cdot n$.  This T-depth is seen by location $\ket{R_{n-1-i}}$ of quantum register $\ket{R}$ (where $0 \leq i \leq n-1$).  We use a quantum \textit{Ctrl-Add} circuit of T-depth $2 \cdot n$ in this Step.

\item Step 4 does not require T gates.

\end{itemize}

We now illustrate the steps we use to determine the total T-depth for the proposed quantum restoring integer division circuit.  

\begin{itemize}	

\item Step 1: Calculate the total T-depth for quantum register $\ket{B}$.  We determine the T-depth for quantum register $\ket{B}$ for each step of the proposed methodology.  

\begin{itemize}

\item Step 1: Locations $\ket{B_{1}}$ through $\ket{B_{n-2}}$ of quantum register $\ket{B}$ see $10$ T gate layers.

\item Step 2 does not require T gates

\item Step 3: Locations $\ket{B_{1}}$ through $\ket{B_{n-2}}$ of quantum register $\ket{B}$ see $13$ T gate layers.

\item Step 4 does not require T gates.

\end{itemize}

Thus, quantum register $\ket{B}$ has a T-depth of $23 \cdot n$.

\item Step 2: Calculate the total T-depth for quantum register $\ket{R}$.  We first consider iterations $1$ through $n-1$.  For iteration $i$ (where $1 \leq i \leq n-1$), we determine the T-depth for quantum register $\ket{R}$ for each step of the proposed methodology. 

\begin{itemize}

\item Step 1: Locations $\ket{R_{n-2-i}}$ through $\ket{R_{0}}$ of quantum register $\ket{R}$ see $4$ T gate layers.  After iteration $n-2$, only location $\ket{R_{0}}$ of quantum register $\ket{R}$ will see T gate layers.   In iteration $n-1$, $\ket{R}$ will not see any T gate layers.

\item Step 2 does not require T gates

\item Step 3: Locations $\ket{R_{n-i}}$ sees $2 \cdot n$ T gate layers while $\ket{R_{n-2-i}}$ through $\ket{R_{0}}$ of quantum register $\ket{R}$ see $6$ T gate layers.  After iteration $n-2$, only locations $\ket{R_{2}}$ and $\ket{R_{0}}$ of quantum register $\ket{R}$ will see T gate layers.   In iteration $n-1$, only location $\ket{R_1}$ will see any T gate layers.  

\item Step 4 does not require T gates.

\end{itemize}

Thus, a single iteration of the proposed design see a T-depth of $2 \cdot n$ on quantum register location $\ket{R_{n-i}}$.  We now consider the final iteration pf the proposed design:

\begin{itemize}

\item Step 1: No locations of quantum register $\ket{R}$ see T gate layers.

\item Step 2 does not require T gates

\item Step 3: Location $\ket{R_{0}}$ sees $2 \cdot n$ T gate layers.  

\item Step 4 does not require T gates.

\end{itemize}

Thus, the final iteration of the proposed design see a T-depth of $2 \cdot n$ on quantum register location $\ket{R_{0}}$.  We calculate the total T-depth seen by each location in register $\ket{R}$ and determined that location $\ket{R_0}$ sees the most T gate layers, a total of $12 \cdot n - 18$ T gate layers.

\item Step 3: Calculate the total T-depth for quantum register $\ket{Q}$.  We first consider iterations $1$ through $n-1$.  For iteration $i$ (where $1 \leq i \leq n-1$), we determine the T-depth for quantum register $\ket{Q}$ for each step of the proposed methodology. 
   
\begin{itemize}

\item Step 1: Locations $\ket{Q_{n-1}}$ through $\ket{Q_{n-i}}$ of quantum register $\ket{Q}$ see $4$ T gate layers.  

\item Step 2 does not require T gates

\item Step 3: Locations $\ket{Q_{n-1}}$ through $\ket{Q_{n-i}}$ of quantum register $\ket{Q}$ see $6$ T gate layers. 

\item Step 4 does not require T gates.

\end{itemize}

Thus, a single iteration of the proposed design see a T-depth of $10$ on quantum register location $\ket{Q_{n-1}}$ through $\ket{Q_{n-i}}$.  We now consider the final iteration of the proposed design:

\begin{itemize}

\item Step 1:  Locations $\ket{Q_{n-2}}$ through $\ket{Q_{0}}$ of quantum register $\ket{Q}$ see $4$ T gate layers.

\item Step 2 does not require T gates

\item Step 3: Locations $\ket{Q_{n-2}}$ through $\ket{Q_{0}}$ of quantum register $\ket{Q}$ see $6$ T gate layers.

\item Step 4 does not require T gates.

\end{itemize}

Thus, the final iteration of the proposed design see a T-depth of $10$ on quantum register locations $\ket{Q_{n-2}}$ through $\ket{Q_{0}}$.  We calculate the total T-depth seen by each location in register $\ket{Q}$ and determined that location $\ket{Q_{n-1}}$ sees the most T gate layers.  Location $\ket{Q_{n-1}}$ sees $10 \cdot n - 8$ T gate layers.

\item Step 4: Determine which quantum register sees the most T gate layers.  In our proposed design the T-depth for each quantum register is given as:

\begin{itemize}

\item Quantum register $\ket{Q}$ has a T-depth of $10 \cdot n - 8$.

\item Quantum register $\ket{R}$ has a T-depth of $12 \cdot n - 18$.

\item Quantum register $\ket{B}$ has a T-depth of $23 \cdot n $.

\end{itemize}

Quantum register $\ket{B}$ sees the most T gate layers because $23 \cdot n > 12 \cdot n - 18$ and $23 \cdot n > 10 \cdot n - 8$.  Thus, the T-depth for our proposed quantum restoring integer division circuit is $23 \cdot n $.  This T-depth is seen by location $\ket{B_{1}}$ through $\ket{B_{n-2}}$ of quantum register $\ket{B}$.      

\end{itemize}    

\subsection{Cost Comparison}

\begin{table}[thbp]
\caption{Comparison of Resource Count Between Proposed and Existing Work }
%\flushleft
%\begin{figure*}[tbh]
\centering

\begin{tabular}{lcccc}
\\ \midrule										
	&		&		1										&		2		&		Proposed		\\ \cmidrule{1-1} \cmidrule{3-5}
T-count	&		&	$	\approx 400 \cdot n^2	$									&	$	\approx 9 \cdot n^3	$	&		$ 35 \cdot n^2 - 28 \cdot n $		\\
T-depth	&		&	$	130 \cdot n	$									&		NA		&	$	23 \cdot n	$	\\
qubits	&		&	$	4 \cdot n	$									&	$	\approx \frac{1}{2} n^3 + 4 \cdot n	$	&		$3 \cdot n$		\\ \bottomrule

\multicolumn{5}{l}{1 is the design by Khosropour et al. \cite{khosropour2011quantum}}\\		
\multicolumn{5}{l}{2 is the design by Dibbo et al. \cite{Dibbo} modified to remove garbage output.}\\
\multicolumn{5}{l}{Table entries are marked NA where a closed-form expression is not} \\
\multicolumn{5}{l}{available for the design by Dibbo et al.}		\\
\end{tabular}

\label{t count comparision table}
%\end{figure*}
\end{table}

\begin{table}[tbhp]
\caption{T-count Comparison of Quantum Integer Division Circuits}
\label{restoringdiv-table:2}
\centering

\resizebox{\columnwidth}{!}{
\begin{tabular}{cccccccc}
\midrule					
$n$	&	&		1	&						2	&	Proposed	&	&		\% Impr.					&	\% Impr.	\\
	&	&			&							&		&	&		w.r.t. 1					&	w.r.t. 2	\\ \cmidrule{1-1} \cmidrule{3-5} \cmidrule{7-8}
4	&	&	$ \approx $	6400	&					$ \approx $	576	&	448	&	&	$ \approx $	93.00					&	$ \approx $ 22.22	\\
8	&	&	$ \approx $	25600	&					$ \approx $	4608	&	2016	&	&	$ \approx $	92.13					&	$ \approx $ 56.25	\\
16	&	&	$ \approx $	102400	&					$ \approx $	36864	&	8512	&	&	$ \approx $	91.69					&	$ \approx $ 76.91	\\
32	&	&	$ \approx $	409600	&					$ \approx $	294912	&	34944	&	&	$ \approx $	91.47					&	$ \approx $ 88.15	\\
64	&	&	$ \approx $	1638400	&					$ \approx $	2359296	&	141568	&	&	$ \approx $	91.36					&	$ \approx $ 94.00	\\
128	&	&	$ \approx $	6553600	&					$ \approx $	18874368	&	569856	&	&	$ \approx $	91.30					&	$ \approx $ 96.98	\\
256	&	&	$ \approx $	26214400	&					$ \approx $	150994944	&	2286592	&	&	$ \approx $	91.28					&	$ \approx $ 98.49	\\
512	&	&	$ \approx $	104857600	&					$ \approx $	1207959552	&	9160704	&	&	$ \approx $	91.26					&	$ \approx $ 99.24	\\ \bottomrule
\multicolumn{5}{r}{Average:}														&	&		91.69					&	79.03	\\ 	\midrule
\multicolumn{8}{l}{1 is the design by Khosropour et al. \cite{khosropour2011quantum}}\\		
\multicolumn{8}{l}{2 is the design by Dibbo et al. \cite{Dibbo} modified to remove garbage output.}\\
\end{tabular}
}
\end{table}

\begin{table}[tbhp]
\caption{Qubit Cost Comparison of Quantum Integer Division Circuits}
\label{restoringdiv-table:3}
\centering
\resizebox{\columnwidth}{!}{
\begin{tabular}{cccccccc}
\midrule
n	&	&	1	&						2	&	Proposed	&	&	\% Impr.	&					\% Impr.	\\
	&	&		&							&		&	&	w.r.t. 1	&					w.r.t. 2	\\ \cmidrule{1-1} \cmidrule{3-5} \cmidrule{7-8}
4	&	&	16	&					$ \approx $	48	&	12	&	&	25.00	&			$ \approx $		75.00	\\
8	&	&	32	&					$ \approx $	288	&	24	&	&	25.00	&			$ \approx $		91.67	\\
16	&	&	64	&					$ \approx $	2112	&	48	&	&	25.00	&		$ \approx $			97.73	\\
32	&	&	128	&					$ \approx $	16512	&	96	&	&	25.00	&		$ \approx $			99.42	\\
64	&	&	256	&					$ \approx $	131328	&	192	&	&	25.00	&		$ \approx $			99.85	\\
128	&	&	512	&					$ \approx $	1049088	&	384	&	&	25.00	&		$ \approx $			99.96	\\
256	&	&	1024	&					$ \approx $	8389632	&	768	&	&	25.00	&	$ \approx $				99.99	\\
512	&	&	2048	&					$ \approx $	67110912	&	1536	&	&	25.00	&		$ \approx $			99.99	\\ \bottomrule 
\multicolumn{5}{r}{Average:}													&	&	25.00	&					93.94	\\  \midrule
\multicolumn{8}{l}{1 is the design by Khosropour et al. \cite{khosropour2011quantum}}\\			
\multicolumn{8}{l}{2 is the design by Dibbo et al. \cite{Dibbo} modified to remove garbage output.}\\

\end{tabular}
}
\end{table}

%%%%%%%%%%%%%%%%%%%%%%%%%%%%%%%%%%%%%%%%%%%%%%%%%%%%%%%%%%%%%%%%%%%%%%%%%%%%%%%%%%%%%%

Comparisons of the proposed design with the current state of the art are presented in Tables \ref{t count comparision table}, \ref{restoringdiv-table:2} and \ref{restoringdiv-table:3}.   We compare our proposed design to the existing quantum restoring division circuit by Khosropour et al. \cite{khosropour2011quantum}.  We also compare against the novel design methodology presented in Dibbo et. al. \cite{Dibbo}.   To perform the comparison of our proposed design to the existing quantum division circuits, we implemented each designs with Clifford+T gates.   To realize reversible gates such as the Toffoli gate, we use the Clifford+T implementations presented in \cite{Maslov}.  As ahown in \cite{Maslov}, the Toffoli gate has a T-count of 7 and a T-depth of 3.  The design by Khosropour et al. depends on quantum gates that cannot be accurately realized with Clifford+T gates.  For these quantum gates, we use the Clifford+T approximations presented in \cite{Kliuchnikov} to calculate the T-count.  We select the approximations for these gates with the poorest accuracy in our calculations since they have the lowest T-count.  We also apply the Bennett's garbage removal scheme illustrated in \cite{Bennett1973trashremoval} to remove the garbage outputs from the design by Dibbo et. al.  We determined the qubit cost for the design in Khosropour et al. by summing the qubits required for the quotient, remainder, garbage outputs, and primary inputs.  We estimated the qubit cost for the design by Dibbo et. al. by summing the qubits required for the quotient, remainder, garbage outputs, and primary inputs.

\subsubsection{Cost Comparison in Terms of T-count}		

Table \ref{t count comparision table} shows that our proposed design and the design by Khosropour et al. have T-count costs of order $\mathcal{O}(n^2)$ while the T-count for the design by Dibbo et al. is of order $\mathcal{O}(n^3)$.	 Table \ref{restoringdiv-table:2} shows that our proposed design methodology achieves improvement ratios ranging from $91.26 \% $ to $93.00 \%$ and $22.22 \%$ to $99.24 \%$ compared to the designs by Khosropour et al. and the design by Dibbo et al. in terms of T-count.  

\subsubsection{Cost Comparison in Terms of Qubits}	

Table \ref{t count comparision table} shows that our proposed design and the design by Khosropour et al. have qubit costs of order $\mathcal{O}(n)$ while the qubit cost for the design by Dibbo et al. is of order $\mathcal{O}(n^3)$.   We also compared the qubit cost of our proposed design methodology to the designs presented by Khosropour et al. and Dibbo et. al. for values of $n$ ranging from $4$ to $512$ in table \ref{restoringdiv-table:3}.  We calculated that our proposed design methodology achieves an improvement ratio of $25.00 \% $ compared to the design by Khosropour et al.  We determined that our proposed design achieves improvement ratios ranging from $75.00 \%$ to $99.99 \% $ compared to the design by Dibbo et. al.

\subsubsection{Cost Comparison in Terms of T-depth}		 

The T-depth cost of the proposed design and designs by Khosropour et al. and the proposed design are $\mathcal{O}(n)$.  A closed-form T-depth expression is not available for the design by Dibbo et al.  We calculated that the design by Khosropour et al. has a T-depth that is $5.6$ times higher than the T-depth of the proposed work.

\section{Proposed Design of Non-Restoring Quantum Integer Division Circuit}
\label{INIS-nonres}

We now present our proposed non-restoring quantum integer division circuit.  The proposed design produces no garbage output and has lower T-count and qubit costs compared to the existing work.  The quantum circuits required to build our proposed quantum non-restoring division circuit are (i) the quantum subtractor, (ii) the quantum \textit{Add-Sub} circuit and (iii) the quantum \textit{Ctrl-Add} circuit.  Our proposed quantum non-restoring divider saves T gates by not doing computation in the QFT domain.  We also base our design on the T gate efficient quantum subtractor, quantum \textit{Add-Sub} circuit and the novel quantum \textit{Ctrl-Add} circuit presented in Section \ref{md_ref}.  The modules used in our proposed quantum circuit do not produce garbage outputs and restore inputs to their original values.  Thus, we are able to save qubits and T gates by placing these quantum circuits such that our proposed quantum restoring division circuit will produce no garbage outputs.  We are able to save additional qubits and T gates because the remainder will be at most $n-1$ bits wide when we divide two $n$ bit numbers with our proposed divider.     

This proposed quantum integer division circuit calculates division by implementing the non-restoring division algorithm.  The non-restoring division algorithm is illustrated in Figure \ref{sqrt-table:70}.  Existing research has demonstrated the correctness of the non-restoring division algorithm through functionally correct circuit implementations such as those in \cite{jamal2013efficient} \cite{mine2017divider}.  %A specific example illustrating how Algorithm 1 calculates the division of a number $a$ by $b$ is available in appendix \ref{sqrt-explain}.

%RESUME HERE!!!!!!!!!!!!!!!!!!!!!!!!!!!!!!!!!!!!!!!!!!!!!!!!!!!!!

%algorithm

\begin{figure}[thbp]

\begin{tabular}{ll}
\\ \midrule
\multicolumn{2}{l}{\textbf{Algorithm 2:} Non-restoring division algorithm} \\ \toprule
\multicolumn{2}{l}{\textbf{Function} Non-Restoring($a,b$) }\\ 
\multicolumn{2}{l}{Requirements: $a$ and $b$ are positive and $2$'s complement. }\\
\multicolumn{2}{l}{\qquad //Takes 2 $n$ bit values $a$ and $b$ as input. }\\
\multicolumn{2}{l}{\qquad //Returns the quotient as an $n$ bit number $Q$ and }\\
\multicolumn{2}{l}{\qquad //the remainder from the division as an $n-1$ bit }\\
\multicolumn{2}{l}{\qquad //number $R$. }\\  
& \\
1 & $R = 0^{n-1}$; // Where $0^{n-1}$ are $n-1$ zeros. \\
2 & $Q = 0^{n-1} a_{n-1}$; // Where $0^{n-1}$ are $n-1$ zeros.\\
3 &\qquad // $Q$'s least significant bit has the value $a_{n-1}$ \\ 
4 &\qquad // $a_{n-1}$ is the most significant bit of $a$. \\
5 & $Q = Q - b$ \\
6 &\\
7 & \textbf{For} $i = 1 \text{ to } n-1$ \\ 
8 & \qquad $Q_{n-i} = \overline{Q_{n-i}}$ \\
9 & \qquad $Y = Q_{n-1-i} \cdots Q_0 R_{n-2} \cdots R_{n-1-i}$ \\
10 & \qquad \qquad // Where $Q_{n-1-i}$ is the most \\
11 & \qquad \qquad // significant bit of $Y$. \\
12 & \qquad \textbf{If} $(Q_{n-i} = 0)$ \\
13 & \qquad \qquad $Y = Y + b$ \\
14 & \qquad \textbf{Else} \\
15 & \qquad \qquad $Y = Y - b$ \\
16 & \qquad \textbf{End} \\
17 & \textbf{End} \\
18 &\\
19 & \textbf{If} $(R < 0)$ \\
20 & \qquad $R = R + b$ \\
21 & \textbf{End} \\
22 & $Q_{0} = \overline{Q_{0}}$ \\
23 &\textbf{Return} $Q, R$ \\ \bottomrule
\end{tabular}

\caption{The non-restoring division algorithm. }
\label{sqrt-table:70}
\end{figure}

%layout done must slim down....

  Consider the division of two $n$ bit $2$'s complement positive binary numbers $a$ and $b$.   Let $\ket{B}$ be a $n$ bit quantum register where that is initialized to the value $b$, let $\ket{R}$ be a $n-1$ quantum register where each register location is initialized with the value $a_i$ for $0 \leq i \leq n-2$ and let $\ket{Q}$ be a $n$ bit quantum register where register location $\ket{Q_0}$ is initialized with the value $a_{n-1}$ and the remaining $n-1$ locations in $\ket{Q}$ are initialized to $0$.  At the end of computation, the quantum register $\ket{B}$ will be restored to the value $b$ while the quantum register $\ket{R}$ will have the remainder of the division of $a$ by $b$.  At the end of computation, the quantum register $\ket{Q}$ will have the quotient of the division of $a$ by $b$.

\begin{figure}[t!bhp]
\centering
\scriptsize

\begin{subfigure}[th]{1.35in}

\[
\Qcircuit @C=0.3em @R=0.5em @!R{
\lstick{\ket{B_{5:0}}}&	&\ctrl{6}			&\qw		&\rstick{\ket{B_{5:0}}} \\
\lstick{\ket{R_0}}&	&\qw			&\qw		&\rstick{\ket{R_0}} \\
\lstick{\ket{R_1}}&	&\qw			&\qw		&\rstick{\ket{R_1}} \\
\lstick{\ket{R_2}}&	&\qw		&\qw		&\rstick{\ket{R_2}} \\
\lstick{\ket{R_3}}&	&\qw		&\qw		&\rstick{\ket{R_3}} \\
\lstick{\ket{R_4}}&	&\qw		&\qw		&\rstick{\ket{R_4}} \\
\lstick{\ket{Q_0}}&	&\multigate{5}{\begin{sideways} Subtraction \end{sideways}}					&\qw		&\rstick{\ket{Q_{0}}} \\
\lstick{\ket{Q_1}}&	&\ghost{\begin{sideways} Subtraction \end{sideways}} &\qw		& \rstick{\ket{Q_1}} \\
\lstick{\ket{Q_2}}&	&\ghost{\begin{sideways} Subtraction \end{sideways}}			&\qw		&\rstick{\ket{Q_2}} \\
\lstick{\ket{Q_3}}&	&\ghost{\begin{sideways} Subtraction \end{sideways}}			&\qw		&\rstick{\ket{Q_3 }} \\
\lstick{\ket{Q_4}}&	&\ghost{\begin{sideways} Subtraction \end{sideways}}			&\qw		&\rstick{\ket{Q_4 }} \\
\lstick{\ket{Q_5}}&	&\ghost{\begin{sideways} Subtraction \end{sideways}}			&\qw		&\rstick{\ket{Q_5 }} \\
}
\]	
\caption{After Step 1}	
\label{s1nonres}	
\end{subfigure} \qquad \begin{subfigure}[th]{1in}
\centering
\scriptsize

\[
\Qcircuit @C=0.3em @R=0.5em @!R{
\lstick{\ket{B_{5:0}}}&	&\ctrl{6}			&\qw &\ctrl{5}	&\qw	&\rstick{\ket{B_{5:0}}} \\
\lstick{\ket{R_0}}&	&\qw			&\qw		&\qw &\qw &\rstick{\ket{R_0}} \\
\lstick{\ket{R_1}}&	&\qw			&\qw		&\qw &\qw &\rstick{\ket{R_1}} \\
\lstick{\ket{R_2}}&	&\qw			&\qw		&\qw &\qw &\rstick{\ket{R_2}} \\
\lstick{\ket{R_3}}&	&\qw			&\qw		&\qw &\qw &\rstick{\ket{R_3}} \\
\lstick{\ket{R_4}}&	&\qw		&\qw		&\multigate{5}{\begin{sideways} Add-Sub \end{sideways}}		&\qw &\rstick{\ket{R_4}} \\
\lstick{\ket{Q_0}}&	&\multigate{5}{\begin{sideways} Subtraction \end{sideways}}					&\qw &\ghost{\begin{sideways} Add-Sub \end{sideways}}		&\qw &\rstick{\ket{Q_{0}}} \\
\lstick{\ket{Q_1}}&	&\ghost{\begin{sideways} Subtraction \end{sideways}} &\qw		&\ghost{\begin{sideways} Add-Sub \end{sideways}} &\qw & \rstick{\ket{Q_1}} \\
\lstick{\ket{Q_2}}&	&\ghost{\begin{sideways} Subtraction \end{sideways}}			&\qw		&\ghost{\begin{sideways} Add-Sub \end{sideways}} &\qw &\rstick{\ket{Q_2}} \\
\lstick{\ket{Q_3}}&	&\ghost{\begin{sideways} Subtraction \end{sideways}}			&\qw		&\ghost{\begin{sideways} Add-Sub \end{sideways}} &\qw &\rstick{\ket{Q_3}} \\
\lstick{\ket{Q_4}}&	&\ghost{\begin{sideways} Subtraction \end{sideways}}			&\qw		&\ghost{\begin{sideways} Add-Sub \end{sideways}} &\qw &\rstick{\ket{Q_4}} \\
\lstick{\ket{Q_5}}&	&\ghost{\begin{sideways} Subtraction \end{sideways}}			&\targ &\ctrl{-1} &\qw		&\rstick{\ket{Q_5 }} \\
}
\]

\caption{After iteration 1 of Step 2}
\label{s21nonres}
\end{subfigure} 
\\ 
\begin{subfigure}[th]{3in}
\centering
\scriptsize

\[
\Qcircuit @C=0.3em @R=0.5em @!R{
\lstick{\ket{B_{5:0}}}&	&\ctrl{6}			&\qw &\ctrl{5}			&\qw &\ctrl{4}			&\qw &\ctrl{3}	&\qw	&\ctrl{2} &\qw &\ctrl{1} &\qw &\rstick{\ket{B_{5:0}}} \\
\lstick{\ket{R_0}}&	&	\qw &	\qw &	\qw &	\qw &\qw	&	\qw 		&\qw		&\qw &\qw &\qw &\multigate{5}{\begin{sideways} Add-Sub \end{sideways}} &\qw &\rstick{\ket{R_0}} \\
\lstick{\ket{R_1}}&	&	\qw &	\qw &	\qw &	\qw &\qw	&	\qw 		&\qw		&\qw &\multigate{5}{\begin{sideways} Add-Sub \end{sideways}} &\qw &\ghost{\begin{sideways} Add-Sub \end{sideways}} &\qw  &\rstick{\ket{R_1}} \\
\lstick{\ket{R_2}}&	&	\qw &	\qw &	\qw  &	\qw &\qw &\qw		&\multigate{5}{\begin{sideways} Add-Sub \end{sideways}}		&\qw &\ghost{\begin{sideways} Add-Sub \end{sideways}} &\qw &\ghost{\begin{sideways} Add-Sub \end{sideways}} &\qw  &\rstick{\ket{R_2}} \\
\lstick{\ket{R_3}}&	&	\qw  &	\qw &	\qw  &	\qw &\multigate{5}{\begin{sideways} Add-Sub \end{sideways}}					&\qw &\ghost{\begin{sideways} Add-Sub \end{sideways}}		&\qw &\ghost{\begin{sideways} Add-Sub \end{sideways}} &\qw &\ghost{\begin{sideways} Add-Sub \end{sideways}} &\qw  &\rstick{\ket{R_{3}}} \\
\lstick{\ket{R_4}}& &	\qw &	\qw &\multigate{5}{\begin{sideways} Add-Sub \end{sideways}}					&\qw &\ghost{\begin{sideways} Add-Sub \end{sideways}}		&\qw &\ghost{\begin{sideways} Add-Sub \end{sideways}} &\qw &\ghost{\begin{sideways} Add-Sub \end{sideways}} &\qw  &\ghost{\begin{sideways} Add-Sub \end{sideways}} &\qw &\rstick{\ket{R_{4}}} \\
\lstick{\ket{Q_0}}&	&\multigate{5}{\begin{sideways} Subtraction \end{sideways}}					&\qw &\ghost{\begin{sideways} Add-Sub \end{sideways}}		&\qw &\ghost{\begin{sideways} Add-Sub \end{sideways}} &\qw &\ghost{\begin{sideways} Add-Sub \end{sideways}} &\qw &\ghost{\begin{sideways} Add-Sub \end{sideways}} &\qw &\ghost{\begin{sideways} Add-Sub \end{sideways}} &\qw  &\rstick{\ket{Q_{0}}} \\
\lstick{\ket{Q_1}}&	&\ghost{\begin{sideways} Subtraction \end{sideways}} &\qw		&\ghost{\begin{sideways} Add-Sub \end{sideways}} &\qw &\ghost{\begin{sideways} Add-Sub \end{sideways}} &\qw &\ghost{\begin{sideways} Add-Sub \end{sideways}} &\qw &\ghost{\begin{sideways} Add-Sub \end{sideways}} &\targ &\ctrl{-1} &\qw & \rstick{\ket{Q_1}} \\
\lstick{\ket{Q_2}}&	&\ghost{\begin{sideways} Subtraction \end{sideways}} &\qw		&\ghost{\begin{sideways} Add-Sub \end{sideways}} &\qw &\ghost{\begin{sideways} Add-Sub \end{sideways}} &\qw &\ghost{\begin{sideways} Add-Sub \end{sideways}} &\targ &\ctrl{-1} &\qw &\qw &\qw & \rstick{\ket{Q_2}} \\
\lstick{\ket{Q_3}}&	&\ghost{\begin{sideways} Subtraction \end{sideways}} &\qw		&\ghost{\begin{sideways} Add-Sub \end{sideways}} &\qw &\ghost{\begin{sideways} Add-Sub \end{sideways}} &\targ &\ctrl{-1} &\qw &\qw &\qw &\qw &\qw  & \rstick{\ket{Q_3}} \\
\lstick{\ket{Q_4}}&	&\ghost{\begin{sideways} Subtraction \end{sideways}}			&\qw		&\ghost{\begin{sideways} Add-Sub \end{sideways}} &\targ &\ctrl{-1} &\qw &\qw &\qw &\qw &\qw &\qw &\qw&\rstick{\ket{Q_4}} \\
\lstick{\ket{Q_5}}&	&\ghost{\begin{sideways} Subtraction \end{sideways}}			&\targ &\ctrl{-1} &\qw		&\qw &\qw &\qw &\qw &\qw &\qw &\qw &\qw &\rstick{\ket{Q_5 }} \\
}
\]

\caption{After final iteration of Step 2}
\label{s2fnonres}
\end{subfigure} 
\\
\begin{subfigure}[th]{3in}
\centering
\scriptsize

\[
\Qcircuit @C=0.2em @R=0.5em @!R{
\lstick{\ket{B_{4:0}}}&	&\ctrl{6}			&\qw &\ctrl{5}			&\qw &\ctrl{4}			&\qw &\ctrl{3}	&\qw	&\ctrl{2} &\qw &\ctrl{1} &\qw &\ctrl{2} &\qw &\qw &\rstick{\ket{B_{4:0}}} \\
\lstick{\ket{B_{5}}}&	&\ctrl{6}			&\qw &\ctrl{5}			&\qw &\ctrl{4}			&\qw &\ctrl{3}	&\qw	&\ctrl{2} &\qw &\ctrl{1} &\qw &\qw &\qw &\qw &\rstick{\ket{B_{5}}} \\
\lstick{\ket{R_0}}&	&	\qw &	\qw &	\qw &	\qw &\qw	&	\qw 		&\qw		&\qw &\qw &\qw &\multigate{5}{\begin{sideways} Add-Sub \end{sideways}} &\qw &\multigate{4}{\begin{sideways} Ctrl-Add \end{sideways}} &\qw &\qw &\rstick{\ket{R_0}} \\
\lstick{\ket{R_1}}&	&	\qw &	\qw &	\qw &	\qw &\qw	&	\qw 		&\qw		&\qw &\multigate{5}{\begin{sideways} Add-Sub \end{sideways}} &\qw &\ghost{\begin{sideways} Add-Sub \end{sideways}} &\qw &\ghost{\begin{sideways} Ctrl-Add \end{sideways}} &\qw  &\qw &\rstick{\ket{R_1}} \\
\lstick{\ket{R_2}}&	&	\qw &	\qw &	\qw  &	\qw &\qw &\qw		&\multigate{5}{\begin{sideways} Add-Sub \end{sideways}}		&\qw &\ghost{\begin{sideways} Add-Sub \end{sideways}} &\qw &\ghost{\begin{sideways} Add-Sub \end{sideways}} &\qw &\ghost{\begin{sideways} Ctrl-Add \end{sideways}} &\qw  &\qw &\rstick{\ket{R_2}} \\
\lstick{\ket{R_3}}&	&	\qw  &	\qw &	\qw  &	\qw &\multigate{5}{\begin{sideways} Add-Sub \end{sideways}}					&\qw &\ghost{\begin{sideways} Add-Sub \end{sideways}}		&\qw &\ghost{\begin{sideways} Add-Sub \end{sideways}} &\qw &\ghost{\begin{sideways} Add-Sub \end{sideways}} &\qw &\ghost{\begin{sideways} Ctrl-Add \end{sideways}} &\qw  &\qw &\rstick{\ket{R_{3}}} \\
\lstick{\ket{R_4}}& &	\qw &	\qw &\multigate{5}{\begin{sideways} Add-Sub \end{sideways}}					&\qw &\ghost{\begin{sideways} Add-Sub \end{sideways}}		&\qw &\ghost{\begin{sideways} Add-Sub \end{sideways}} &\qw &\ghost{\begin{sideways} Add-Sub \end{sideways}} &\qw  &\ghost{\begin{sideways} Add-Sub \end{sideways}} &\qw &\ghost{\begin{sideways} Ctrl-Add \end{sideways}} &\qw &\qw &\rstick{\ket{R_{4}}} \\
\lstick{\ket{Q_0}}&	&\multigate{5}{\begin{sideways} Subtraction \end{sideways}}					&\qw &\ghost{\begin{sideways} Add-Sub \end{sideways}}		&\qw &\ghost{\begin{sideways} Add-Sub \end{sideways}} &\qw &\ghost{\begin{sideways} Add-Sub \end{sideways}} &\qw &\ghost{\begin{sideways} Add-Sub \end{sideways}} &\qw &\ghost{\begin{sideways} Add-Sub \end{sideways}} &\qw  &\ctrl{-1} &\targ &\qw &\rstick{\ket{Q_{0}}} \\
\lstick{\ket{Q_1}}&	&\ghost{\begin{sideways} Subtraction \end{sideways}} &\qw		&\ghost{\begin{sideways} Add-Sub \end{sideways}} &\qw &\ghost{\begin{sideways} Add-Sub \end{sideways}} &\qw &\ghost{\begin{sideways} Add-Sub \end{sideways}} &\qw &\ghost{\begin{sideways} Add-Sub \end{sideways}} &\targ &\ctrl{-1} &\qw &\qw &\qw &\qw & \rstick{\ket{Q_1}} \\
\lstick{\ket{Q_2}}&	&\ghost{\begin{sideways} Subtraction \end{sideways}} &\qw		&\ghost{\begin{sideways} Add-Sub \end{sideways}} &\qw &\ghost{\begin{sideways} Add-Sub \end{sideways}} &\qw &\ghost{\begin{sideways} Add-Sub \end{sideways}} &\targ &\ctrl{-1} &\qw &\qw &\qw &\qw &\qw &\qw & \rstick{\ket{Q_2}} \\
\lstick{\ket{Q_3}}&	&\ghost{\begin{sideways} Subtraction \end{sideways}} &\qw		&\ghost{\begin{sideways} Add-Sub \end{sideways}} &\qw &\ghost{\begin{sideways} Add-Sub \end{sideways}} &\targ &\ctrl{-1} &\qw &\qw &\qw &\qw &\qw  &\qw &\qw &\qw & \rstick{\ket{Q_3}} \\
\lstick{\ket{Q_4}}&	&\ghost{\begin{sideways} Subtraction \end{sideways}}			&\qw		&\ghost{\begin{sideways} Add-Sub \end{sideways}} &\targ &\ctrl{-1} &\qw &\qw &\qw &\qw &\qw &\qw &\qw &\qw &\qw &\qw &\rstick{\ket{Q_4}} \\
\lstick{\ket{Q_5}}&	&\ghost{\begin{sideways} Subtraction \end{sideways}}			&\targ &\ctrl{-1} &\qw		&\qw &\qw &\qw &\qw &\qw &\qw &\qw &\qw &\qw &\qw &\qw &\rstick{\ket{Q_5 }} \\
}
\]

\caption{After Step 3}
\label{s3nonres}
\end{subfigure}

\caption{Circuit generation of proposed quantum non-restoring division circuit (illustrated with a 6 bit divider).}
\label{div:FIG1}
\end{figure}
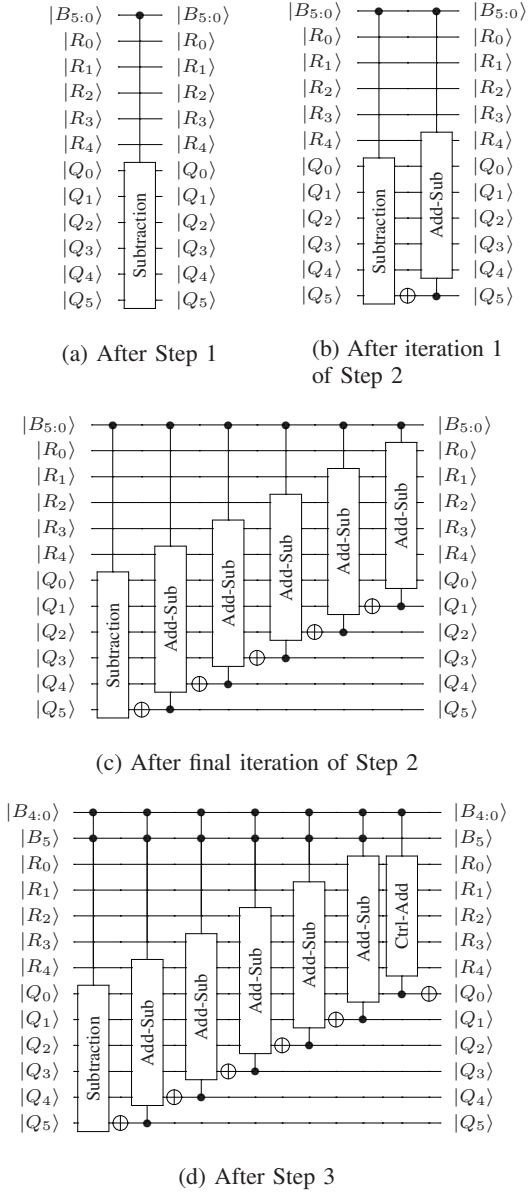

The proposed methodology is generic in nature and can design a quantum non-restoring integer division circuit of any size.  The steps of the proposed methodology are presented along with an illustrative example of the proposed quantum non-restoring integer division circuit for the division of two $6$ bit numbers $a_0 \dots a_5$ and $b_0 \dots b_5$ shown in Figure \ref{div:FIG1}.  The proposed methodology has three steps.  A quantum circuit is generated for each step of the design.

%%%%%%%%%%%%%%%%%%%%%%%%%%%%%%%%%%%%%%%%%%%%%%%%%%%%%%%%%%%%%%%%%%%%%%%%%%%%%%%%%%%%%%%%%%%%%%%%%%%%%%%%%%%%%%%%%%%%%%%%% 

\subsection{Steps of the Proposed Design Methodology}

\begin{itemize}

\item Step 1: This step executes line 5 of Algorithm 2 in quantum hardware.  Step 1 is shown for a 6 bit restoring divider in Figure \ref{s1nonres}.  Apply the quantum registers $\ket{Q}$ and $\ket{B}$ to a quantum subtraction circuit such that, at the end of computation, the quantum register $\ket{B}$ is unchanged while quantum register $\ket{Q}$ now has the result of computation.  

\item Step 2: this step is repeated $n-1$ times and has the following four sub-steps.    This step executes lines 8 through 16 of Algorithm 2 in quantum hardware.  Iteration 1 of Step 2 is shown for a 6 bit restoring divider in Figure \ref{s21nonres}.  The final Iteration of Step 2 is shown for a 6 bit restoring divider in Figure \ref{s2fnonres}.  We show the steps for iteration $i$ where $ 1 \leq i \leq n-1$.    

\begin{itemize}

\item Sub-step 1:  This sub-step executes line 9 of Algorithm 2 in quantum hardware.  Treat the locations $\ket{R_{n-2}}$ through $\ket{R_{n-1-i}}$ of quantum register $\ket{R}$ and locations $\ket{Q_{n-1-i}}$ through $\ket{Q_{0}}$ of register $\ket{Q}$ as a combined quantum register $\ket{Y}$.  The values at locations $\ket{R_{n-2}}$ through $\ket{R_{n-1-i}}$ will occupy locations $\ket{Y_{i-1}}$ through $\ket{Y_0}$ and the values at locations $\ket{Q_{n-1-i}}$ to $\ket{Q_{0}}$ will occupy locations $\ket{Y_{n-1}}$ through $\ket{Y_{i}}$. 

\item Sub-step 2: This sub-step executes line 8 of Algorithm 2 in quantum hardware and prepares location $\ket{Q_{n-i}}$ for use in subsequent sub-steps..  At quantum register location $\ket{Q_{n-i}}$ apply a quantum NOT gate.  Location $\ket{Q_{n-i}}$ now has the quotient bit $q_{n-i}$ of the division of $a$ by $b$.     

\item Sub-step 3: Apply the quantum registers $\ket{B}$ and $\ket{Y}$ to a quantum \textit{Add-Sub} circuit such that $\ket{B}$ is unchanged while $\ket{Y}$ will hold the result of computation.    

\item Sub-step 4: Apply the quantum register location $\ket{Q_{n-i}}$ to the quantum \textit{Add-Sub} circuit such that the operation of the circuit is conditioned on the value at location $\ket{Q_{n-i}}$.  Location $\ket{Q_{n-i}}$ is unchanged.  

\end{itemize}

If $\ket{Q_{n-i}} = 0$, this step executes line 13 of Algorithm 2.  If $\ket{Q_{n-i}} = 1$, this step executes line 15 of Algorithm 2.

\item Step 3: This step executes lines 19 through 22 of Algorithm 2 in quantum hardware.  Step 3 is shown for a 6 bit restoring divider in Figure \ref{s3nonres}.  This step has the following three sub-steps:

\begin{itemize}

\item Sub-step 1: Apply locations $\ket{B_{n-2}}$ through $\ket{B_{0}}$ of quantum register $\ket{B}$ and quantum register $\ket{R}$ to a quantum \textit{Ctrl-Add} circuit such that locations $\ket{B_{n-2}}$ through $\ket{B_{0}}$ are unchanged while quantum register $\ket{R}$ is transformed to the remainder.  

\item Sub-step 2: Apply location $\ket{Q_{0}}$ of register $\ket{Q}$ to the quantum \textit{Ctrl-Add} circuit such that the operation of the circuit is conditioned on the value at register location $\ket{Q_{0}}$.  Location $\ket{Q_{0}}$ is unchanged.  If $\ket{Q_{0}} = 1$, Step 1 and Step 2 execute line 21.  After this Sub-step 2, quantum register $\ket{R}$ will contain the remainder of the division of $a$ by $b$.    

\item Sub-step 3: At quantum register location $\ket{Q_{0}}$, apply a quantum NOT gate.  Step 3 executes line 23 of Algorithm 2.  Location $\ket{Q_{0}}$ now has the quotient bit $q_{0}$ of the division of $a$ by $b$.  After this step, quantum register $\ket{Q}$ will contain the quotient of the division of $a$ by $b$.  

\end{itemize}

\end{itemize}

%%%%%%%%%%%%%%%%%%%%%%%%%%%%%%%%%%%%%%%%%%%%%%%%%%%%%%%%%%%%%%%%%%%%%%%%%%%%%%%%%%%%%%%%%%%%%%%%%%%%%%%%%%%%%%%%%%%%%%%%%%%%%%%%%
% there is a purpose for this madness  RESUME HERE

\begin{table*}[t!bhp]
\caption{Comparison of Quantum Integer Division Circuits}
\label{nrsdiv-table:1}
\centering
%\resizebox{\columnwidth}{!}{
\begin{tabular}{cccccc}
 \midrule
 & & 1 & 2 & 3 & Proposed \\ \cmidrule{1-1} \cmidrule{3-6}
T count	&		&					$	28 \cdot n^2 	$	&	$	42 \cdot n^2 + 28 \cdot n	$	&	$	\approx 9 \cdot n^3	$		&	$	14 \cdot n^2 + 7 \cdot n + 7	$	\\
T-depth & 		& NA	 &	NA  &NA &	$10 \cdot n + 13$ \\						
qubits	&		&					$	2 \cdot n^2 + 5 \cdot n -1	$	&	$	3 \cdot n^2 + 14 \cdot n	$	&	$	\approx \frac{1}{2} n^3 + 4 \cdot n	$		&		$3 \cdot n -1$		\\

 \bottomrule
\multicolumn{6}{l}{1 and 2 are the designs by Jamal et al. \cite{jamal2013efficient} modified to remove garbage output.} \\
\multicolumn{6}{l}{3 is the design by Dibbo et al. \cite{Dibbo} modified to remove garbage output.}\\
\multicolumn{6}{l}{Table entries are marked NA where a closed-form expression is not available for } \\
\multicolumn{6}{l}{the designs by Jamal et al. and Dibbo et al.}		\\
\end{tabular}
\end{table*}

\begin{table*}[tbhp]
\caption{T-count Comparison of Quantum Integer Division Circuits}
\label{nrsdiv-table:2}
\centering

%\resizebox{\columnwidth}{!}{
\begin{tabular}{cccccccccc}
\midrule					
$n$	& &					1	&	2	&		3	&	Proposed	&	&					\% Impr.	&	\% Impr.	&		\% Impr.	\\
	& &						&		&			&		&	&					w.r.t. 1	&	w.r.t. 2	&		w.r.t. 3	\\ \cmidrule{1-1} \cmidrule{3-6} \cmidrule{8-10}
4	& &					448	&	784	&	$ \approx $	576	&	217	&	&					51.56	&	72.32	&	$ \approx $	62.33	\\
8	& &					1792	&	2912	&	$ \approx $	4608	&	917	&	&					48.83	&	68.51	&	$ \approx $	80.10	\\
16	& &					7168	&	11200	&	$ \approx $	36864	&	3661	&	&					48.93	&	67.31	&	$ \approx $	90.07	\\
32	& &					28672	&	43904	&	$ \approx $	294912	&	14525	&	&					49.34	&	66.92	&	$ \approx $	95.07	\\
64	& &					114688	&	173824	&	$ \approx $	2359296	&	57757	&	&					49.64	&	66.77	&	$ \approx $	97.55	\\
128	& &					458752	&	691712	&	$ \approx $	18874368	&	230237	&	&					49.81	&	66.71	&	$ \approx $	98.78	\\
256	& &					1835008	&	2759680	&	$ \approx $	150994944	&	919261	&	&					49.90	&	66.69	&	$ \approx $	99.39	\\
512	& &					7340032	&	11024384	&	$ \approx $	1207959552	&	3673565	&	&					49.95	&	66.68	&	$ \approx $	99.70	\\ \bottomrule
\multicolumn{6}{r}{Average:}															&	&				49.75	&	67.74	&		90.37	\\ \midrule
\multicolumn{10}{l}{1 and 2 are the designs by Jamal et al. \cite{jamal2013efficient} modified to remove garbage output.}	\\	
\multicolumn{10}{l}{3 is the design by Dibbo et al. \cite{Dibbo} modified to remove garbage output.}\\

\end{tabular}
\end{table*}

%%%%%%%%%%%%%%%%%%%%%%%%%%%%%%%%%%%%%%%%%%%%%%%%%%%%%%%%%%%%%%%%%%%%%%%%%%%%%%%%%%%%%%%%%%%%%%%%%%%%%%%%%%%%%%%%%%%%%%%%%%%%%%%%%%%%%%%%

\section{Cost Analysis of the Proposed Non-Restoring Division Circuit}
\label{div-nrs-cost}

\begin{table*}[tbhp]
\caption{Qubit Cost Comparison of Quantum Integer Division Circuits}
\label{nrsdiv-table:3}
\centering
%\resizebox{\columnwidth}{!}{
\begin{tabular}{cccccccccc}
\midrule
$n$	& &			1	&	2	&		3	&	Proposed	&	&			\% Impr.	&	\% Impr.	&	\% Impr.	\\
	& &				&		&			&		&	&			w.r.t. 1	&	w.r.t. 2	&	w.r.t. 3	\\ \cmidrule{1-1} \cmidrule{3-6} \cmidrule{8-10}
4	& &			51	&	104	&	$ \approx $	48	&	11	&	&			78.43	&	89.42	&	$ \approx $ 77.08	\\
8	& &			167	&	304	&	$ \approx $	288	&	23	&	&			86.23	&	92.43	&	$ \approx $ 92.01	\\
16	& &			591	&	992	&	$ \approx $	2112	&	47	&	&			92.05	&	95.26	&	$ \approx $ 97.77	\\
32	& &			2207	&	3520	&	$ \approx $	16512	&	95	&	&			95.70	&	97.30	&	$ \approx $ 99.42	\\
64	& &			8511	&	13184	&	$ \approx $	131328	&	191	&	&			97.76	&	98.55	&	$ \approx $ 99.85	\\
128	& &			33407	&	50944	&	$ \approx $	1049088	&	383	&	&			98.85	&	99.25	&	$ \approx $ 99.96	\\
256	& &			132351	&	200192	&	$ \approx $	8389632	&	767	&	&			99.42	&	99.62	&	$ \approx $ 99.99	\\
512	& &			526847	&	793600	&	$ \approx $	67110912	&	1535	&	&			99.71	&	99.81	&	$ \approx $ 99.99	\\ \bottomrule
\multicolumn{6}{r}{Average:}												&	&			93.52	&	96.46	&	95.76	\\ \midrule
\multicolumn{10}{l}{1 and 2 are the two designs from Jamal et al. \cite{jamal2013efficient} modified to remove garbage output.}	\\	
\multicolumn{10}{l}{3 is the design by Dibbo et al. \cite{Dibbo} modified to remove garbage output.}\\
\end{tabular}
\end{table*}

\subsection{T-Count Analysis}

The T-count of the proposed quantum integer division circuit is illustrated shortly for each step of the proposed design methodology:

\begin{itemize}

\item The T-count for Step 1 is $14 \cdot n - 14$.  We use a quantum subtraction circuit of T-count $14 \cdot n - 14$ in this step.

\item Step 2 is repeated $n-1$ times.  The T-count for each iteration of Step 2 is $14 \cdot n - 14$.  We use a quantum \textit{Add-Sub} circuit of T-count $14 \cdot n - 14$ in this step.

\item The T-count for Step 3 is $21 \cdot n - 21$.  We use a quantum \textit{Ctrl-Add} circuit of size $n-1$ in this step.  We use a quantum \textit{Ctrl-Add} circuit of T-count $21 \cdot n -14$ in this step.

\end{itemize}

We determine the total T-count by summing the T-count for each step in the design as shown below:

\begin{equation}
14 \cdot n - 14 + (14 \cdot n - 14) \cdot (n-1) + 21 \cdot n -21
\label{nrsdiv-EQ:1}
\end{equation}

This expression can be simplified to the following:

\begin{equation}
14 \cdot n^2 + 7 \cdot n - 35
\label{nrsdiv-EQ:2}
\end{equation}

\subsection{T-depth Cost}

The T-depth of the proposed quantum integer division circuit is illustrated shortly for each step of the design methodology.  Our proposed design is based on T-depth efficient designs of quantum subtractors, quantum \textit{Add-Sub} circuits and quantum \textit{Ctrl-Add} circuits.  We determined that garbageless and T gate optimized quantum subtractor and quantum \textit{Add-Sub} circuits in the literature such as the designs in \cite{Thapliyal2016addsub} have a T-depth that is constant and independent of the circuit size $n$.  Thus, these quantum circuits have T-depth of order $\mathcal{O}(1)$.  We determined as well that \textit{Ctrl-Add} circuits in the literature such as the design in \cite{edgard2017multiplier} scale as a function of circuit size $n$.  Thus, these \textit{Ctrl-Add} circuits have a T-depth of order $\mathcal{O}(n)$.  
		
\begin{itemize}
			
\item Step 1 has a constant T-depth of $10$.  This T-depth is seen by locations $\ket{B_{1}}$ through $\ket{B_{n-2}}$ of quantum register $\ket{B}$.  We use a quantum subtraction circuit of constant T-depth $10$ in this step (where T-depth is independent of circuit size $n$). 
			
\item Step 2 is repeated $n-1$ times.   The $i$th iteration of Step 2 (where $1 \leq i \leq n-1$) has a constant T-depth of $10$.  This T-depth is seen by locations $\ket{B_{1}}$ through $\ket{B_{n-2}}$ of quantum register $\ket{B}$.  We use a quantum \textit{Add-Sub} circuit of constant T-depth $10$ in this step (where T-depth is independent of circuit size $n$). 
			
\item Step 3 has a T-depth of $2 \cdot n$.  This T-depth is seen by location $\ket{Q_0}$ of quantum register $\ket{Q}$.  We use a quantum \textit{Ctrl-Add} circuit of T-depth $2 \cdot n$ in this Step.

\end{itemize}	
		
We now illustrate the steps to determine the total T-depth for the proposed quantum integer division circuit:

\begin{enumerate}

\item Calculate the T-depth for Step 1.  Step 1 has a T-depth of $10$.  This T-depth is seen by locations $\ket{B_{1}}$ through $\ket{B_{n-2}}$ of quantum register $\ket{B}$.

\item Calculate the T-depth for all iterations of Step 2.  The total T-depth for all iteration of Step 2 has a T-depth of $10 \cdot (n-1)$ because each iteration of Step 2 requires a quantum \textit{Add-Sub} circuit.  The total T-depth $10 \cdot (n-1 )$ simplifies to $10 \cdot n - 10$.   This T-depth is seen by locations $\ket{B_{1}}$ through $\ket{B_{n-2}}$ of quantum register $\ket{B}$.

\item Calculate the T-depth for Step 3.  Step 3 has a T-depth of $2 \cdot n$.  This T-depth is seen by location $\ket{Q_0}$ of quantum register $\ket{Q}$.

\item Determine which qubits see the most T gate layers.  We find after comparing all the qubits in our proposed design, quantum register location $\ket{Q_0}$ of quantum register $\ket{Q}$ and quantum register locations $\ket{B_{1}}$ through $\ket{B_{n-2}}$ of quantum register $\ket{B}$ see the most T gate layers.

\item Determine the total number of T gate layers seen by quantum register location $\ket{Q_0}$ in the proposed design.  Quantum register $\ket{Q_0}$ will see a total of $6 \cdot n-4$ T gate layers because in Step 1 location $\ket{Q_0}$ sees $4$ T gate layers, in Step 2 location $\ket{Q_0}$ sees $4 \cdot (n-2)$ T gate layers and in Step 2 location $\ket{Q_0}$ sees $2 \cdot n$ T gate layers.  The total number of T gate layers seen by location $\ket{Q_0}$ is $4 + 4 \cdot (n-2) + 2 \cdot n$ which simplifies to $6 \cdot n-4$.

\item Determine the total number of T gate layers seen by quantum register locations $\ket{B_{1}}$ through $\ket{B_{n-2}}$ in the proposed design.  Quantum register locations $\ket{B_{1}}$ through $\ket{B_{n-2}}$ will see a total of $10 \cdot n+13$ T gate layers because in Step 1 locations $\ket{B_{1}}$ through $\ket{B_{n-2}}$ see $10$ T gate layers, in Step 2 locations $\ket{B_{1}}$ through $\ket{B_{n-2}}$ sees $10 \cdot (n-1)$ T gate layers and in Step 3 locations $\ket{B_{1}}$ through $\ket{B_{n-2}}$ see $13$ T gate layers.  The total number of T gate layers seen by locations $\ket{B_{1}}$ through $\ket{B_{n-2}}$ is $10 + 10 \cdot (n-1) + 13$ which simplifies to $10 \cdot n+13$.

\item Determine which qubits see the most T gate layers.  We determined that locations $\ket{B_{1}}$ through $\ket{B_{n-2}}$ see more T gate layers than register location $\ket{Q_0}$ because $10 \cdot n+13 > 6 \cdot n -4$.  

\end{enumerate}    

Thus, our proposed design has a T-depth of $10 \cdot n + 13$ and this T-depth is seen by locations $\ket{B_{1}}$ through $\ket{B_{n-2}}$ of quantum register $\ket{B}$.

\subsection{Cost Comparison}

Comparison of the proposed design with the current state of the art are presented in Tables \ref{nrsdiv-table:1}, \ref{nrsdiv-table:2} and \ref{nrsdiv-table:3}.  We compare our proposed design to the existing quantum non-restoring division circuits by Jamal et al. \cite{jamal2013efficient} and the alternative design methodology presented in Dibbo et. al. \cite{Dibbo}.  To perform the comparison we implemented each design with Clifford+T gates.  To realize reversible gates such as the Toffoli gate, we use the Clifford+T implementations presented in \cite{Maslov}.  As ahown in \cite{Maslov}, the Toffoli gate has a T-count of 7 and a T-depth of 3.  We also apply the Bennett's garbage removal scheme illustrated in \cite{Bennett1973trashremoval} to remove the garbage outputs from each design presented by Jamal et al. and Dibbo et. al.  We determined the qubit cost for each design in Jamal et. al. by summing the qubits required for the quotient, remainder, garbage outputs, and primary inputs.  We estimated the qubit cost for each design by summing the qubits required for the quotient, remainder, garbage outputs, and primary inputs.

\subsubsection{Cost Comparison in Terms of T-count}		

Table \ref{nrsdiv-table:1} illustrates that the T-count cost of the proposed design and the designs by Jamal et al. are $\mathcal{O}(n^2)$.  The design by Dibbo et. al. has a T-count cost of order $\mathcal{O}(n^3)$.   Table \ref{nrsdiv-table:2} shows that our proposed design methodology achieves improvement ratios ranging from $49.95 \%$ to $51.56 \% $, $66.68 \%$ to $72.32 \%$ and $62.33 \%$ to $99.70 \%$ compared to the designs by Jamal et al. and the design by Dibbo et. al. in terms of T-count.  

\subsubsection{Cost Comparison in Terms of Qubits}	

Table \ref{nrsdiv-table:1} shows that our proposed design has a qubit cost of order $\mathcal{O}(n)$ while the qubit cost for the designs by Jamal et al. are of order $\mathcal{O}(n^2)$.  Table \ref{nrsdiv-table:1} also illustrates that the design by Dibbo et. al. has a qubit cost of order $\mathcal{O}(n^3)$.  Table \ref{nrsdiv-table:3} shows the comparison of our proposed design methodology to the designs presented by Jamal et al. and Dibbo et. al. for values of $n$ ranging from $4$ to $512$ in terms of qubit cost.  We calculated that our proposed design methodology achieves improvement ratios ranging from $78.43 \%$ to $99.71 \%$, $89.42 \%$ to $99.81 \%$ and $77.08 \%$ to $99.99 \%$ compared to the designs by Jamal et al. and the design by Dibbo et. al.		  
	 
\subsubsection{Cost Comparison in Terms of T-depth}		 

The T-depth cost of the proposed design is $\mathcal{O}(n)$.  A closed-form expression is not available for the designs by Jamal et al. and the design by Dibbo et al. for the T-depth.

\section{Conclusion}

In this work, we have proposed two designs for quantum circuit integer division based on Clifford+T gates. The first quantum integer division circuit proposed is based on the restoring division algorithm and the second is based on the non-restoring division algorithm.  We also show the design of components used in our proposed quantum integer division circuits such as the quantum subtraction circuit, quantum \textit{Add-Sub} circuit and quantum \textit{Ctrl-Add} circuit.  The proposed quantum restoring division circuit is shown to be superior to existing designs in terms of T-depth, T-count and qubits.  Likewise, the proposed quantum non-restoring division circuit is shown to be superior to existing designs in terms of T-count and qubits.  We conclude that the proposed restoring division circuit or proposed non-restoring division circuit can be integrated in a larger quantum data path system design where T-count and T-depth are of primary concern. 

\bibliographystyle{IEEEtran}
%\bibliography{ref2.bib}
\bibliography{MULTbiblio-smooth.bib}

\begin{thebibliography}{10}
\providecommand{\url}[1]{#1}
\csname url@samestyle\endcsname
\providecommand{\newblock}{\relax}
\providecommand{\bibinfo}[2]{#2}
\providecommand{\BIBentrySTDinterwordspacing}{\spaceskip=0pt\relax}
\providecommand{\BIBentryALTinterwordstretchfactor}{4}
\providecommand{\BIBentryALTinterwordspacing}{\spaceskip=\fontdimen2\font plus
\BIBentryALTinterwordstretchfactor\fontdimen3\font minus
  \fontdimen4\font\relax}
\providecommand{\BIBforeignlanguage}[2]{{%
\expandafter\ifx\csname l@#1\endcsname\relax
\typeout{** WARNING: IEEEtran.bst: No hyphenation pattern has been}%
\typeout{** loaded for the language `#1'. Using the pattern for}%
\typeout{** the default language instead.}%
\else
\language=\csname l@#1\endcsname
\fi
#2}}
\providecommand{\BIBdecl}{\relax}
\BIBdecl

\bibitem{Quipper}
P.~Selinger{ et al.}, \emph{The Quipper System}, 2016, available at:
  http://www.mathstat.dal.ca/~selinger/quipper/doc/.

\bibitem{Hallgren2017encryption}
\BIBentryALTinterwordspacing
S.~Hallgren, ``Polynomial-time quantum algorithms for pell's equation and the
  principal ideal problem,'' \emph{J. ACM}, vol.~54, no.~1, pp. 4:1--4:19, Mar.
  2007. [Online]. Available: \url{http://doi.acm.org/10.1145/1206035.1206039}
\BIBentrySTDinterwordspacing

\bibitem{vandam2000quadraticcharacter}
\BIBentryALTinterwordspacing
W.~{van Dam} and S.~{Hallgren}, ``{Efficient Quantum Algorithms for Shifted
  Quadratic Character Problems},'' \emph{eprint arXiv}, Nov. 2000. [Online].
  Available: \url{https://arxiv.org/abs/quant-ph/0011067}
\BIBentrySTDinterwordspacing

\bibitem{vandam2006hiddenshift}
\BIBentryALTinterwordspacing
W.~van Dam, S.~Hallgren, and L.~Ip, ``Quantum algorithms for some hidden shift
  problems,'' \emph{SIAM Journal on Computing}, vol.~36, no.~3, pp. 763--778,
  2006. [Online]. Available: \url{https://doi.org/10.1137/S009753970343141X}
\BIBentrySTDinterwordspacing

\bibitem{LIQUi}
D.~Wecker{ et al.}, \emph{Language-Integrated Quantum Operations:
  LIQUi$\vert\rangle$}, 2016, available at:
  https://www.microsoft.com/en-us/research/project/language-integrated-quantum-operations-liqui/.

\bibitem{Fredkin}
\BIBentryALTinterwordspacing
E.~Fredkin and T.~Toffoli, ``Conservative logic,'' \emph{International Journal
  of Theoretical Physics}, vol.~21, no.~3, pp. 219--253, 1982. [Online].
  Available: \url{http://dx.doi.org/10.1007/BF01857727}
\BIBentrySTDinterwordspacing

\bibitem{Zhou-T}
\BIBentryALTinterwordspacing
X.~Zhou, D.~W. Leung, and I.~L. Chuang, ``Methodology for quantum logic gate
  construction,'' \emph{Phys. Rev. A}, vol.~62, p. 052316, Oct 2000. [Online].
  Available: \url{https://link.aps.org/doi/10.1103/PhysRevA.62.052316}
\BIBentrySTDinterwordspacing

\bibitem{Maslov}
M.~Amy, D.~Maslov, M.~Mosca, and M.~Roetteler, ``A meet-in-the-middle algorithm
  for fast synthesis of depth-optimal quantum circuits,'' \emph{IEEE
  Transactions on Computer-Aided Design of Integrated Circuits and Systems},
  vol.~32, no.~6, pp. 818--830, June 2013.

\bibitem{PalerIOP}
\BIBentryALTinterwordspacing
A.~Paler, I.~Polian, K.~Nemoto, and S.~J. Devitt, ``Fault-tolerant, high-level
  quantum circuits: form, compilation and description,'' \emph{Quantum Science
  and Technology}, vol.~2, no.~2, p. 025003, 2017. [Online]. Available:
  \url{http://stacks.iop.org/2058-9565/2/i=2/a=025003}
\BIBentrySTDinterwordspacing

\bibitem{Gosset}
\BIBentryALTinterwordspacing
D.~Gosset, V.~Kliuchnikov, M.~Mosca, and V.~Russo, ``An algorithm for the
  t-count,'' \emph{Quantum Information {\&} Computation}, vol.~14, no. 15-16,
  pp. 1261--1276, 2014. [Online]. Available:
  \url{http://www.rintonpress.com/xxqic14/qic-14-1516/1261-1276.pdf}
\BIBentrySTDinterwordspacing

\bibitem{khosropour2011quantum}
A.~Khosropour, H.~Aghababa, and B.~Forouzandeh, ``Quantum division circuit
  based on restoring division algorithm,'' in \emph{Information Technology: New
  Generations (ITNG), 2011 Eighth International Conference on}.\hskip 1em plus
  0.5em minus 0.4em\relax IEEE, 2011, pp. 1037--1040.

\bibitem{Jain2015quantumdivider}
L.~Jamal and H.~M.~H. Babu, ``Efficient approaches to design a reversible
  floating point divider,'' in \emph{2013 IEEE International Symposium on
  Circuits and Systems (ISCAS2013)}, May 2013, pp. 3004--3007.

\bibitem{Dibbo}
S.~V. Dibbo, H.~M.~H. Babu, and L.~Jamal, ``An efficient design technique of a
  quantum divider circuit,'' in \emph{2016 IEEE International Symposium on
  Circuits and Systems (ISCAS)}, May 2016, pp. 2102--2105.

\bibitem{Kliuchnikov}
\BIBentryALTinterwordspacing
V.~Kliuchnikov, D.~Maslov, and M.~Mosca, ``Fast and efficient exact synthesis
  of single-qubit unitaries generated by clifford and t gates,'' \emph{Quantum
  Info. Comput.}, vol.~13, no. 7-8, pp. 607--630, Jul. 2013. [Online].
  Available: \url{http://dl.acm.org/citation.cfm?id=2535649.2535653}
\BIBentrySTDinterwordspacing

\bibitem{Varun2016divider}
H.~Thapliyal, T.~S.~S. Varun, and E.~Mu{\~{n}}oz-Coreas, ``Quantum circuit
  design of integer division optimizing ancillary qubits and t-count,'' in
  \emph{16th Asian Quantum Information Science Conference}, August 2016, pp.
  197--199.

\bibitem{mine2017divider}
H.~Thapliyal, T.~S.~S. Varun, E.~Mu{\~{n}}oz-Coreas, K.~A. Britt, and T.~S.
  Humble, ``Quantum circuit designs of integer division optimizing t-count and
  t-depth,'' in \emph{2017 IEEE International Symposium on Nanoelectronic and
  Information Systems (iNIS)}, Dec 2017, pp. 123--128.

\bibitem{IBM_quantum}
IBM, \emph{Quantum Computing - IBM Q}, 2017, available at:
  https://www.research.ibm.com/ibm-q/.

\bibitem{Song2017computersaresmall}
C.~{Song}, K.~{Xu}, W.~{Liu}, C.-p. {Yang}, S.-B. {Zheng}, H.~{Deng}, Q.~{Xie},
  K.~{Huang}, Q.~{Guo}, L.~{Zhang}, P.~{Zhang}, D.~{Xu}, D.~{Zheng}, X.~{Zhu},
  H.~{Wang}, Y.-A. {Chen}, C.-Y. {Lu}, S.~{Han}, and J.-W. {Pan}, ``{10-Qubit
  Entanglement and Parallel Logic Operations with a Superconducting Circuit},''
  \emph{Physical Review Letters}, vol. 119, no.~18, p. 180511, Nov. 2017.

\bibitem{Thapliyal2016addsub}
H.~Thapliyal, ``Mapping of subtractor and adder-subtractor circuits on
  reversible quantum gates,'' in \emph{Transactions on Computational Science
  XXVII}.\hskip 1em plus 0.5em minus 0.4em\relax Springer, 2016, pp. 10--34.

\bibitem{thapliyal2013design}
H.~Thapliyal and N.~Ranganathan, ``Design of efficient reversible logic-based
  binary and bcd adder circuits,'' \emph{ACM Journal on Emerging Technologies
  in Computing Systems (JETC)}, vol.~9, no.~3, p.~17, 2013.

\bibitem{edgard2017multiplier}
\BIBentryALTinterwordspacing
E.~{Mu{\~n}oz-Coreas} and H.~{Thapliyal}, ``{T-count Optimized Design of
  Quantum Integer Multiplication},'' \emph{ArXiv e-prints}, Jun. 2017.
  [Online]. Available: \url{https://arxiv.org/abs/1706.05113}
\BIBentrySTDinterwordspacing

\bibitem{Bennett1973trashremoval}
\BIBentryALTinterwordspacing
C.~H. Bennett, ``Logical reversibility of computation,'' \emph{IBM J. Res.
  Dev.}, vol.~17, no.~6, pp. 525--532, Nov. 1973. [Online]. Available:
  \url{http://dx.doi.org/10.1147/rd.176.0525}
\BIBentrySTDinterwordspacing

\bibitem{jamal2013efficient}
L.~Jamal and H.~M.~H. Babu, ``Efficient approaches to design a reversible
  floating point divider,'' in \emph{2013 IEEE International Symposium on
  Circuits and Systems (ISCAS2013)}, May 2013, pp. 3004--3007.

\end{thebibliography}

\end{document}